\documentclass{aa}  

\usepackage{txfonts}
\usepackage{natbib}

\usepackage{microtype}
\usepackage{amssymb,amsmath}
\usepackage{graphicx}
\usepackage{xcolor}
\usepackage{multirow}
\usepackage[T1]{fontenc}
\usepackage{ae,aecompl}
\usepackage{newtxtext,newtxmath}
\usepackage{longtable,listings}
\usepackage{hyperref}

\def\obj{SDSS J1004+1510}

\newcommand{\shorttitle}[1]{%
	\renewcommand{\titlerunning}{#1}%
}

\authorrunning{Liao et al.}

\shorttitle{Optical QPOs in SDSS J1004+1510}

\begin{document}

\title{Optical QPOs with dual periodicities 1103days and 243days in the blue quasar SDSS J100438.8+151056}
	

\author{GuiLin Liao
\inst{1}
\and
XingQian Chen\inst{1}\and Qi Zheng\inst{1}\and YiLin Liu \inst{1}\and XueGuang Zhang\inst{1}\fnmsep\thanks{E-mail: {xgzhang@gxu.edu.cn}}}
	
	\institute{School of Physical Science and Technology, GuangXi University, No.100, Daxue East Road,
		 Nanning, 530004, P. R. China}

	\date{}
	
	
\abstract{
This manuscript investigates the possible existence of a binary supermassive black holes (BSMBH) system in the blue quasar 
SDSS J100438.8+151056 (=SDSS J1004+1510) at z=0.219 based on the detection of robust optical QPOs. We determine QPOs using multiple 
analysis methods applied to the CSS-V, ZTF g/r band light curves, and additionally, combined with the characteristics of broad emission 
lines and explores potential mechanisms for the QPOs, including jet and disk precession models. Two distinct periodicities, $1103\pm260$days 
and $243\pm29$days, are identified in the ZTF g/r-band light curves with confidence level exceeding $5\sigma$, through four different 
techniques. Meanwhile, the $1103\pm260$days periodicity is also clearly detected in the CSS V-band light curve. The optical periodicities 
suggest a BSMBH system candidate in SDSS J1004+1510, with an estimated total virial BH mass of $(1.13\pm0.14)\times10^8 M_{\odot}$ 
and a space separation of $0.0053\pm0.0016$pc for the periodicity of $1103\pm260$days. The second periodicity of $243\pm29$days could 
be attributed to harmonic oscillations, considering $(1103\pm260)/(243\pm29)\sim4.54\pm0.47$ with large scatters. However, if the 
periodicity of $243\pm29$days was from an independent QPO, a triple BH system candidate on sub-pc scale could be probably expected, with 
space separations of $0.00036\pm0.00004$pc between a close BSMBH system and of $0.0053\pm0.0016$pc between the BSMBH 
system and the third BH, after considering similar BH mass of the third BH as the total mass of the central BSMBH. These 
findings strongly demonstrate that combined light curves from the different sky survey projects can lead to more reliable QPOs candidates 
to be detected, and also indicate higher quality light curves could be helpful to find probably potential QPOs with multiple periodicities 
leading to rare detections of candidates for sub-pc triple BH systems.}

	
	\maketitle
	%
	
\section{Introduction}
	
	Galaxy merging is an important part of the formation and evolution of galaxies in the universe \citep{ca92, kw93, lc94, bh96, sr98, 
mh01, lk04, bf09, md06, se14, rp16, rs17, bh19, mj21, jk21, yp22, le23, kk24}. Almost every galaxy has a BH at its center \citep{kr95, ff05, 
kh13, hb14}. It means that the process of galaxy mergers is often accompanied by the formation of dual Active Galactic Nuclei (dual AGN) on 
kpc scale and binary supermassive black holes (BSMBH) systems on sub-pc scale \citep{bb80, mk10, fg19, yg19, mj22}. When 
galaxies merge, the BHs at their centers, due to dynamical friction \citep{mm05, cy20}, gradually move closer to each other and orbit 
around a common center of mass, forming a dual AGN on kpc-scale. As time goes on, gravitational wave radiation becomes the main mechanism 
driving the BHs further closer together, and eventually the distance between them shrinks to on sub-pc scale, forming a tightly packed 
BSMBH system. These two BHs may further merge to form a larger supermassive BH (SMBH), releasing strong gravitational waves 
at the same time \citep{er76, fc94, cb10, hu21}. The study of dual AGN and BSMBHs is one of the hotspots of current research 
and is crucial for understanding the co-evolution of SMBHs and galaxies.

	Dual AGN systems on kpc scales have been widely observed and studied through directly resolved images. However, sub-pc scale  
BSMBH systems have rarely been detected through directly resolved images, and the extremely close proximity between the 
BHs in BSMBH systems makes observation and identification very difficult through photometric images. Even with the help of 
high-resolution techniques such as the Very Long Baseline Interferometry Array (VLBA), only a few successes have been reported. For example, 
\citet{rt06} discovered a BSMBH system within the radio galaxy 0402+379. This discovery was made possible through the use of 
VLBA, which provided high-resolution imaging of two compact, variable, and active nuclei within the galaxy, these nuclei with a projected 
separation of about 7.3 pc. In addition, \citet{kl17} also reported another sub-pc scale BSMBH candidate with a separation 
of approximately 0.35 pc in NGC 7674 through VLBA.

	Given the limitations of direct imaging observations, scientists have also attempted to use a variety of indirect observational 
methods to explore the sub-pc scale BSMBH systems in recent years. Among these, spectroscopic characterization is an 
important tool. For example, by observing the double-peaked features of broad emission lines in the spectra of quasars, \citet{lb09, 
ss09, td11, p94, sl13, ls14, re15, wg17} have analyzed the velocity variations of these broad emission lines. Their findings provide 
evidence supporting the probable existence of sub-pc scale BSMBH systems.

	Besides the applications of unique spectroscopic features of broad emission lines, the detection of QPOs in long-term light 
curves is also an important and convenient indicator for identifying sub-pc scale BSMBHs candidates. Many research teams 
have reported the QPOs in AGN with periodicity ranging from hundreds to thousands of days. These QPOs can be caused by a variety of 
physical mechanisms, including jet procession \citep{mg85, ck92, ab00, ca13, hy21}, BH tidal disruption events \citep{sz20}, accretion 
disk instabilities \citep{vs98, tl09, pv13}, general relativistic effects \citep{sv98, sv99, iv16}, and orbital motions of BSMBH 
systems \citep{kf09, ga10, bb15, sx20}. Although the origin of QPOs in light curves is complex, most reported QPOs are considered to 
be from the orbital motions of BSMBH systems. For example, \citet{gd15a} detected and reported a 5.2yr periodicity in the 
quasar PG 1302-102, attributed to the orbital motion of a BSMBH system. In a follow-up study, \citet{gd15b} conducted a 
systematic search for QPOs in light curves of 243500 sources from the Catalina Real-Time Transient Survey (CRTS). This search led to 
the identification of 111 potential BSMBH candidates. The light curve characteristics of these candidates, as validated 
by theoretical models, were found to be consistent with the periodic variations expected from BSMBH systems. Similarly, 
\citet{lg15, gd15a, gd15b, cb16, lw16, zb16, ky20, ss20, lc21, ok22} have also reported sub-pc scale BSMBH candidates based on QPOs 
detected in the light curves. Additionally, in our recent studies on QPOs, \citet{zh22a, zh22b, zh23b, zh25} have also identified 
such BSMBH candidates in broad line AGN.

	QPOs play a key role in detecting BSMBH systems in galaxies, however, relying on QPO signals for detections can 
present some accuracy challenges. First, the time durations of the light curves are monitored as a key factor, because the detection of 
QPO signals needs to rely on long-term time-series data to ensure that the observed periodic variations are not chance events. In 
addition, the stochastic AGN variability may produce signals similar to QPOs, as discussed in \citet{sh18} and \citet{vu16}. To improve 
the accuracy of detections of QPOs and to reduce false-positive results, it is crucial to select light curves with long-term and 
continuous observational data. Among the public sky surveys, the Catalina Sky Survey (CSS) \citep{md11, dd09} has sufficient temporal 
baseline and sky coverage to provide favorable conditions for searching for long-lived optical QPOs. In addition, the high-frequency 
observing capability of the Zwicky Transient Telescope (ZTF) \citep{bk19, ds20} provides a finer light curve for capturing optical QPOs 
with shorter periods. Combining the long-term light curves of the CSS with the ZTF allows for more efficient identifications and 
analysis of optical QPOs. As in the recent work of \citet{zh22b, zh25}, the author combined the observations of CSS with ZTF at different 
time periods to obtain light-variation data up to 16 years, which provides important clues for detecting more reliable optical QPOs.

	In this manuscript, through the combined light curves from CSS and ZTF, we report the discovery of a potential BSMBH 
system in the blue quasar SDSS J100438.8+151056 (=\obj) (redshift of 0.219), which is not reported in the sample of \citet{gd15b}, probably 
due to short time duration of the CSS V-band light curve of \obj~ relative to its periodicity. After analyzing the CSS V-band light 
curve, an optical QPO signal with a periodicity around 1000 days can be detected. Moreover, QPOs with a periodicity around 220 days 
can be detected in the higher quality ZTF g/r-band light curves, besides the periodicity around 1000 days. The structure of this 
manuscript is as follows. Section 2 presents the optical QPOs through four analytical methods applied to the long-term optical light 
curves of SDSS J1004+1510. Section 3 focuses on the spectroscopic results. Section 4 discusses the probable central BSMBH 
system and the possibility of dual-periodic signals probably related to a triple BH system candidate. Finally, the conclusions are 
given in Section 5. In the manuscript, the cosmological parameters have been adopted as $H_{0}=70{\rm km\cdot s}^{-1}{\rm Mpc}^{-1}$, 
$\Omega_{\Lambda}=0.7$ and $\Omega_{\rm m}=0.3$.
	
	
\section{Optical QPOs in \obj}
\subsection{Long-term optical light curves of \obj}
	
	The long-term light curves in observer frame of SDSS J1004+1510 can be collected from the CSS 
(\href{http://nesssi.cacr.caltech.edu/DataRelease/}{http://nesssi.cacr.caltech.edu/DataRelease/}) and ZTF 
(\href{https://www.ztf.caltech.edu}{https://www.ztf.caltech.edu}). The CSS V-band light curve is collected with MJD-53000 from 469 
(March 2005) to 3592 (November 2013). And, the ZTF g/r band light curves are collected with MJD-53000 from the 5202 (March 2018) to 
6996 (March 2023). Here, due to the ZTF i-band light curve only including 53 data points in a short time duration, it is not considered 
in this manuscript. The collected light curves are shown in the top left panel of Fig.~\ref{mcmc}.

\subsection{The analyses of QPOs in \obj}
	
	Similar to what we have recently done in \citet{zh22b, zh23b, zh25}, the following four methods are applied to determine the 
probable QPOs in \obj, the direct fitting method, the phase-folded method, the Generalized Lomb-Scargle (GLS) method \citep{lo76, sc82, br01, 
zk09, vt18, se20} and the weighted wavelet Z-transform (WWZ) method \citep{fo96, an16, gt18, ks20, lc21}.

\begin{figure*}
\centering\includegraphics[width = 8cm,height=6cm]{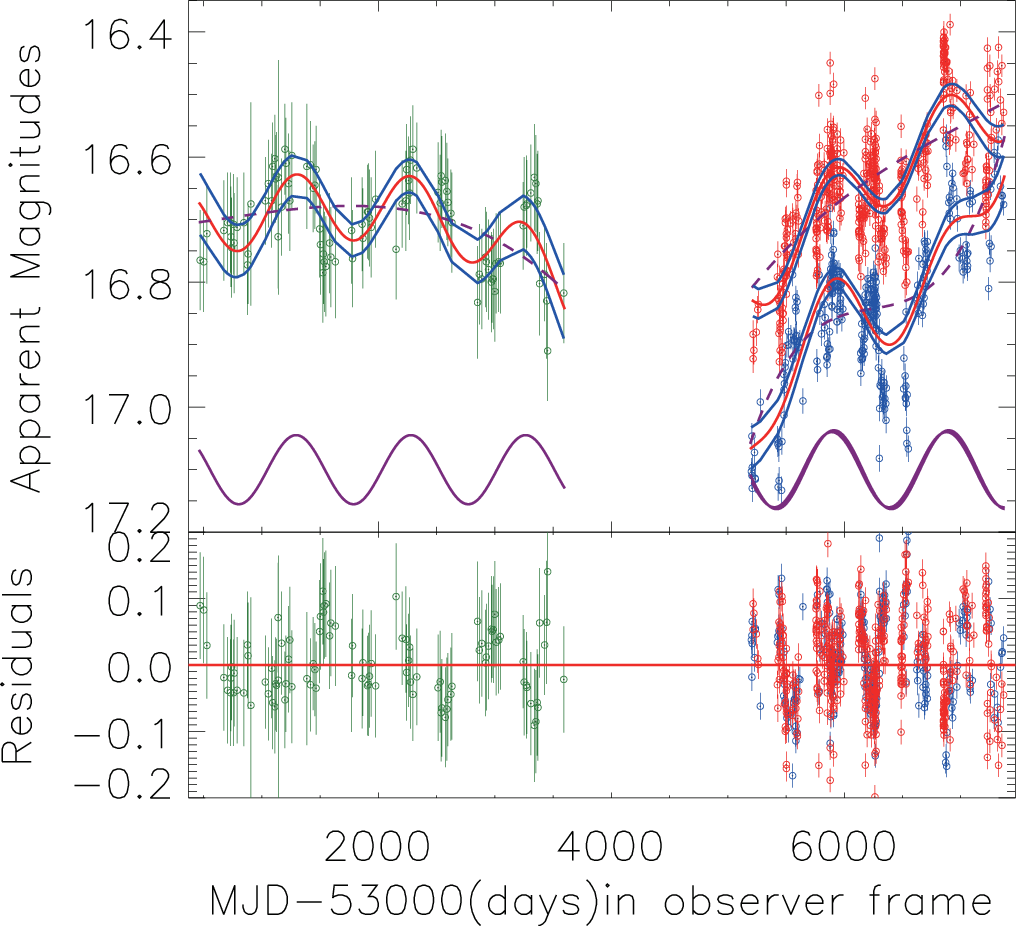}
\centering\includegraphics[width = 8cm,height=6cm]{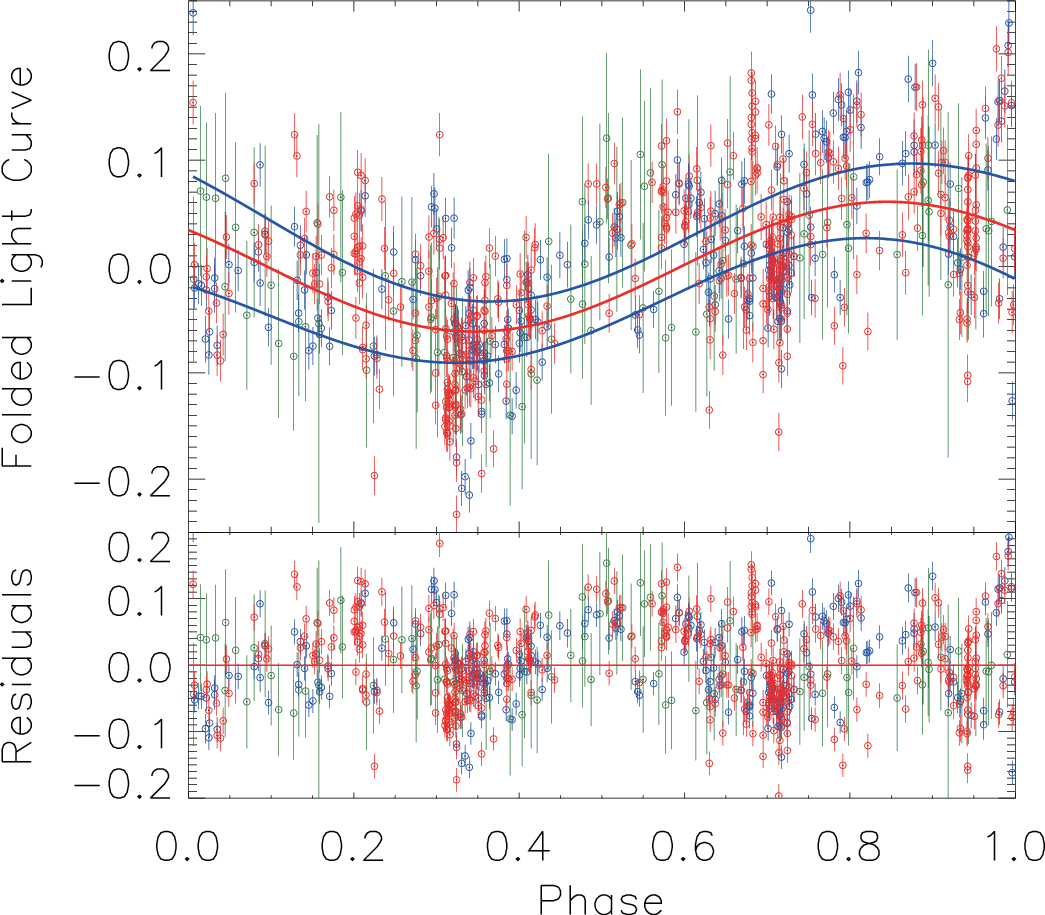}
\caption{The top left panel displays the light curves from the CSS (solid circles plus error bars in green) and from the ZTF in 
the g-band (minus 0.35mag) (open circles plus error bars in blue) and r-band (minus 0.35mag) (open circles plus error bars in red), 
alongside the best-fitting results (solid lines in red) from a sinusoidal function plus a fourth-degree polynomial component. In 
the top left panel, the solid purple lines represent the component described by the sinusoidal function (plus 17.5 mag), while the 
dashed purple lines indicate the component modeled by the polynomial function. The top right panel shows the phase-folded light 
curves based on the determined periodicity 985 days (with the polynomial trends subtracted; green for CSS data, blue for ZTF g-band, 
and red for ZTF r-band) along with the best sinusoidal fit (solid line in red). The solid blue lines in the top panels indicate 
the corresponding 5$\sigma$ confidence bands to the best fitting results through the F-test technique. The bottom panels display 
the corresponding residuals (light curves minus the best fitting results), with the solid red line denoting the residuals=0}
\label{mcmc}
\end{figure*}

	The light curves are first described by the direct fitting method, with a sinusoidal function plus a fourth-degree polynomial 
component simultaneously applied to each light curve from the CSS and the ZTF. The sinusoidal function is specifically chosen to capture 
the periodic variations in the light curve, without delving into the physical origins of the QPOs, while the fourth-degree polynomial 
is used to represent long-term non-periodic trends. The model functions are described as follows:
\begin{equation}
\begin{split}
LMC{1} &= A + B \times (\frac{t}{1000 \text{ days}}) + C \times (\frac{t}{1000 \text{ days}})^2\\ 
& + D \times (\frac{t}{1000 \text{ days}})^3 + E \times \sin(\frac{2\pi t}{T_{QPO}} + \phi_{0})
\end{split}
\label{e1}
\end{equation}
While the model functions are applied, the parameter of $T_{QPO}$ is the same for each light curve. To ensure the best fitting, the 
Levenberg-Marquardt least-squares method is used to optimize the parameters of these composite models. The determined model parameters 
are presented in Table~\ref{table1}, with the determined periodicity to be approximately 985$\pm$8days. The best fitting results and 
the corresponding 99.99994\% (5\(\sigma\)) confidence bands (determined by the F-test technique) are shown in the top left panel of 
Figure \ref{mcmc}, with \(\chi^2/\text{dof} = 8847.144/827\sim10.69\). The residuals (the differences between the observed data and 
the model predictions) are shown in the bottom left panel of Figure \ref{mcmc}.

	Besides the model functions in Equation (1), if only the fourth-degree polynomial components are used to describe each light 
curve, it will lead to \(\chi_1^2/\text{dof}_1 = 13117.669/834\sim15.7\). To assess the significance of the sinusoidal components, 
the F-test technique is employed to compare models with and without the sinusoidal component. The F-test statistic is calculated as follows:
\begin{equation}
F_P = \frac{\frac{x_1^2 - x^2}{\text{dof}_1 - \text{dof}}}{x^2 / \text{dof}} = 57.03
\end{equation}
Based on the given confidence level (5\(\sigma\)), the numerator degree of freedom (\(\text{dof}_1 - \text{dof}\)), and the denominator 
degree of freedom (\(\text{dof}\)), the corresponding F-distribution value is calculated as:
\begin{equation}
F_{\text{dis}} = f_{\text{cvf}}(1 - 99.99994\%, 7, 827) = 6.3
\end{equation}
Therefore, through the F-test statistical technique, the sinusoidal component is preferred with a confidence level much higher than 
$5\sigma$, indicating that the periodic fluctuations in the data prefer to be a genuine feature rather than a random occurrence.

	Based on the periodicity obtained from the direct fitting method, the phase-folded method is applied to the light curves 
after subtracting the polynomial components. Through a simple sinusoidal function by 
\begin{equation}
LMC_{\text{ph}} = A \times \sin(2\pi t + B)+C
\end{equation} 
the folded light curves can be well described and shown in the top right panel of Figure \ref{mcmc}, with 
\(\chi^2/\text{dof} = 8860.0165/843 = 10.51\), through the Levenberg-Marquardt least-squares minimization technique. The bottom 
right panel of Figure \ref{mcmc} shows the residuals (differences between the fitting results and the phase-folded light curves).

	To further investigate the optical QPOs in \obj, in addition to the direct fitting method and the phase-folded method shown 
in Figure \ref{mcmc}, the improved GLS method (see the python astroML package) is also applied. Compared with the traditional 
Lomb-Scargle method, GLS is particularly suitable for handling unevenly sampled datasets, allowing for more accurate identification 
and quantification of periodic signals in time series.

\begin{figure}
\centering\includegraphics[width = 8cm,height=5cm]{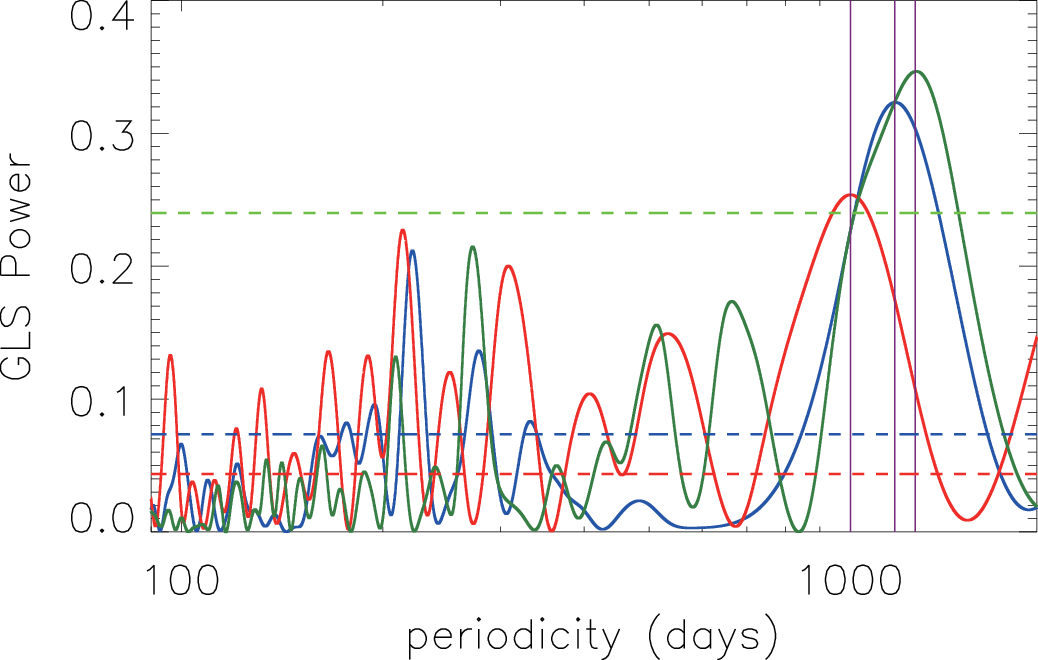}
\caption{Properties of the GLS periodogram. Solid lines in red, blue, and green show the corresponding results from the ZTF r-band, 
g-band, and CSS V-band light curves, respectively. The horizontal dashed lines in red, blue, and green indicate the corresponding 
5\(\sigma\) confidence levels (FAP = 1-99.99994\%) for the results from the ZTF r-band, g-band, and CSS V-band light curves, respectively.}
\label{gls}
\end{figure}

\begin{figure*}
\centering\includegraphics[width = 5cm,height=4cm]{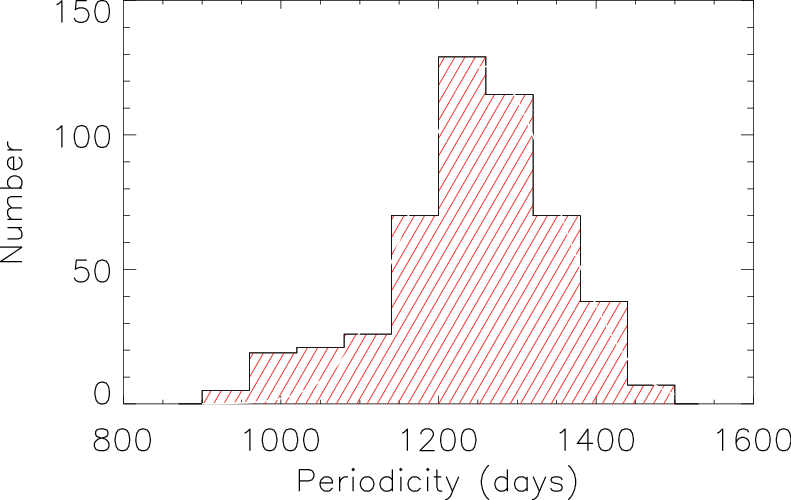}
\centering\includegraphics[width = 5cm,height=4cm]{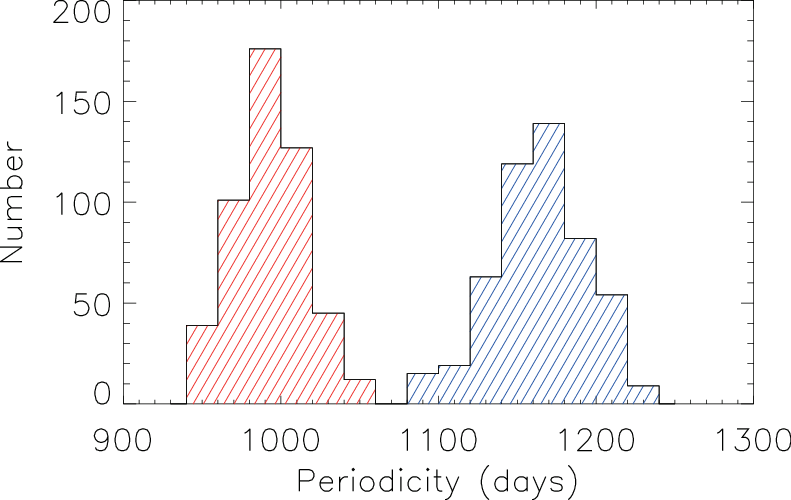}
\centering\includegraphics[width = 5cm,height=4cm]{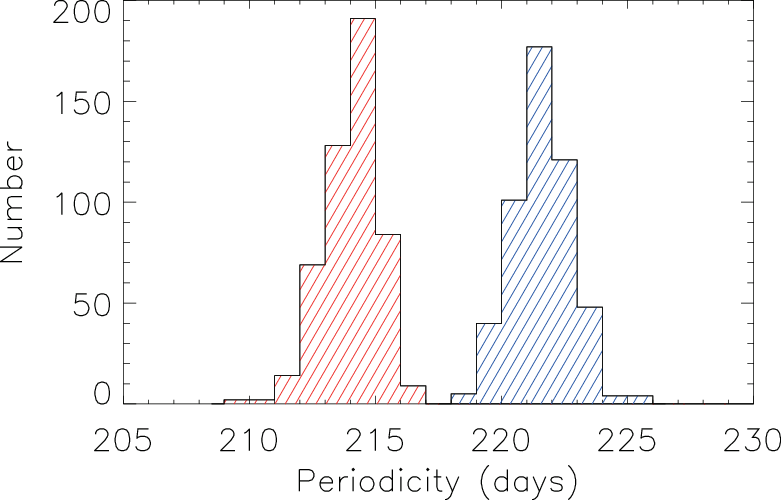}
\caption{On the periodicity distributions through the bootstrap method. The left panel shows the results for the CSS-V band light 
curve. The middle panel and the right panel show the periodicities for the ZTF g-band(blue histogram) around 1165days and 221days, 
as well as for the ZTF r-band(red histogram) around 990days and 215days.}
\label{bstoop}
\end{figure*}

	Figure \ref{gls} presents the GLS periodogram results for the CSS V-band and the ZTF g-band and r-band light curves. In the 
CSS V-band light curve, a significant peak is detected, corresponding to a periodicity about $1250\pm112$days (solid line in green), 
with a significance level higher than 99.99994\%, along with a secondary, less prominent peak around $272\pm1$days (confidence level 
smaller than 5$\sigma$). In the GLS method results for the high quality ZTF g-band and r-band (represented by blue and red lines) 
light curves, there are main peaks around $1165\pm30$days and $1000\pm30$days with significance levels higher than 99.99994\%. 
Interestingly, compared with the CSS V-band light curve, there are more prominent periodicities around $221\pm2$days and $215\pm1$days, 
respectively, both exceeding the 99.99994\% confidence level. Based on the GLS periodogram applied to randomly created light curves 
including about half of the data points in the origin light curves, Figure~\ref{bstoop} shows the distributions of peak periodicity 
in the CSS V-band and ZTF g-band and r-band light curves determined through the bootstrap method applied with 500 loops, leading to 
the determined uncertainties of the periodicities by the half width at half maximum of the distributions.

\begin{figure}
\centering\includegraphics[width = 8cm,height=8cm]{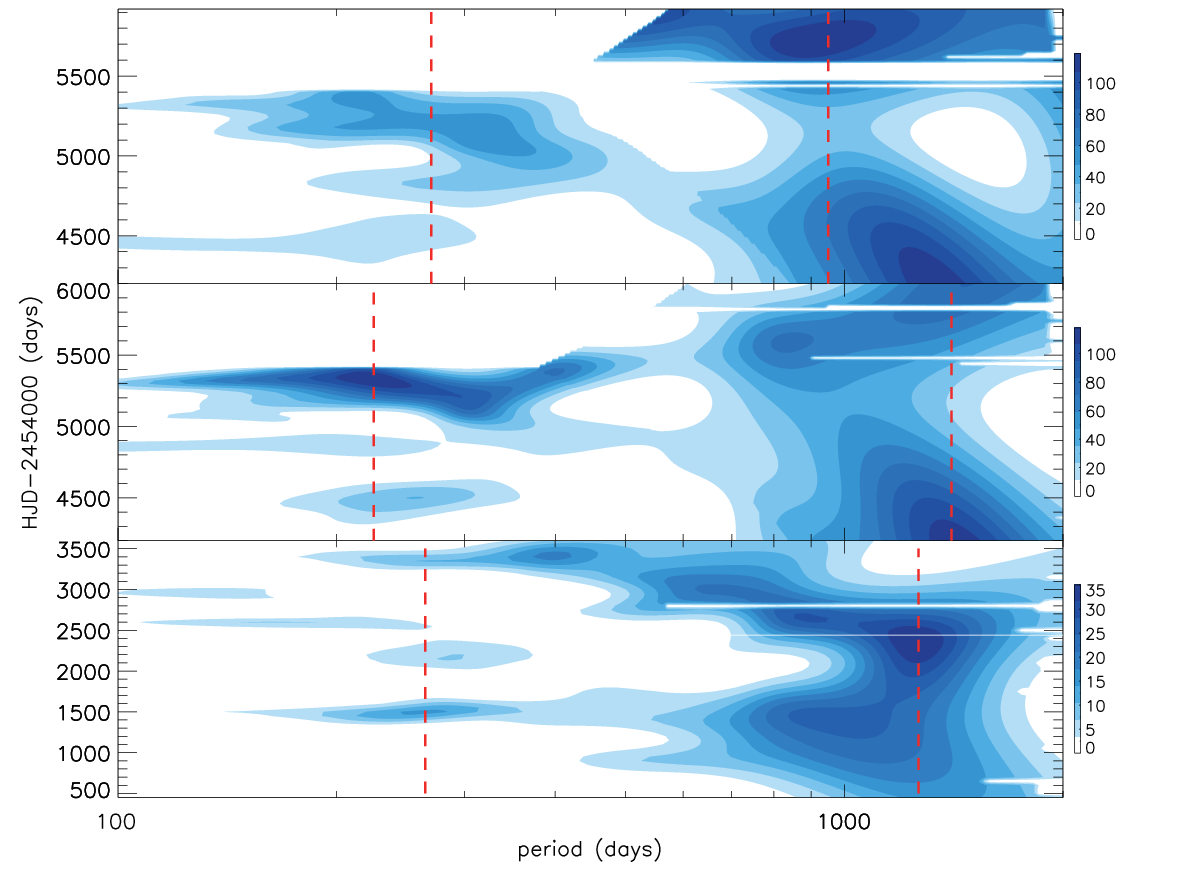}
\caption{The WWZ results for the CSS-V band and the ZTF g/r-band light curves. For the ZTF r-band light curve (top panel), two 
periodicities are identified at approximately $950\pm42$days and $270\pm3.5$days. For the ZTF g-band light curve (middle panel), 
two periodicities are around $1405\pm40$days and $225\pm3$days. For the CSS-V band light curve (bottom panel), two periodicities are 
around $1265\pm40$days and $265\pm3$days. In each panel, the vertical  dashed red lines mark the positions for the periodicities.}
\label{wwz}
\end{figure}

	In addition, the WWZ method is also employed for an in-depth analysis of the CSS V-band and ZTF g/r-band light curves. As 
shown in the Figure \ref{wwz}, the power spectrum covered a frequency (in units of 1/days) range from 0.0005 to 0.01, with a frequency 
step of 0.00001. The result for the ZTF-r band light curve in the top panel reveals two distinct periodicities, one around $950\pm42$days, 
and another around $270\pm3.5$days. In the middle panel from the ZTF g-band light curve, two significant periodicities are also 
observed: one around $1405\pm40$days, and a shorter one around  $225\pm3$days. While the bottom panel from the CSS-V band light curve 
shows a dominant periodicity around $1265\pm40$days, with a secondary periodicity around $265\pm3$days. The results obtained from the 
WWZ method are consistent with those derived from the GLS periodogram.

\begin{figure}
\centering\includegraphics[width = 8cm,height = 3.5cm]{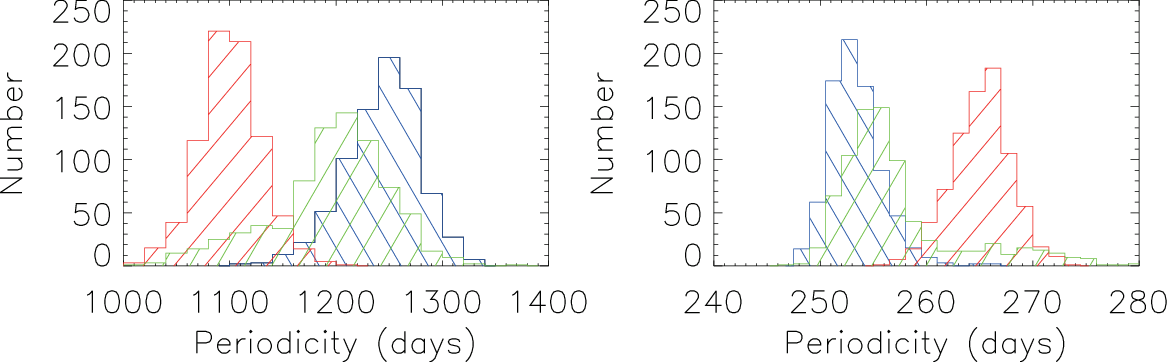}
\caption{The left and right panels present the distributions of WWZ determined periodicities around 1000days and 200days, determined 
by the bootstrap method. The green, blue and red histograms show the results for the CSS V-band and the ZTF g/r band light curves, 
respectively.}
\label{testwwz}
\end{figure}

	Additionally, to determine the uncertainties of the periodicities detected by the WWZ method, the bootstrap technique is 
applied. More than half of the data points from the observed CSS-V and ZTF g/r band light curves are randomly collected to generate 
new light curves, with the process repeated 800 times.  Subsequently, the WWZ method is used to determine the periodic distributions 
of these new light curves, and the results are shown in Figure \ref{testwwz}. The uncertainties of the periodicities were determined 
by the half width at half maximum of the distributions.

\begin{figure}
\centering\includegraphics[width = 8cm,height=6cm]{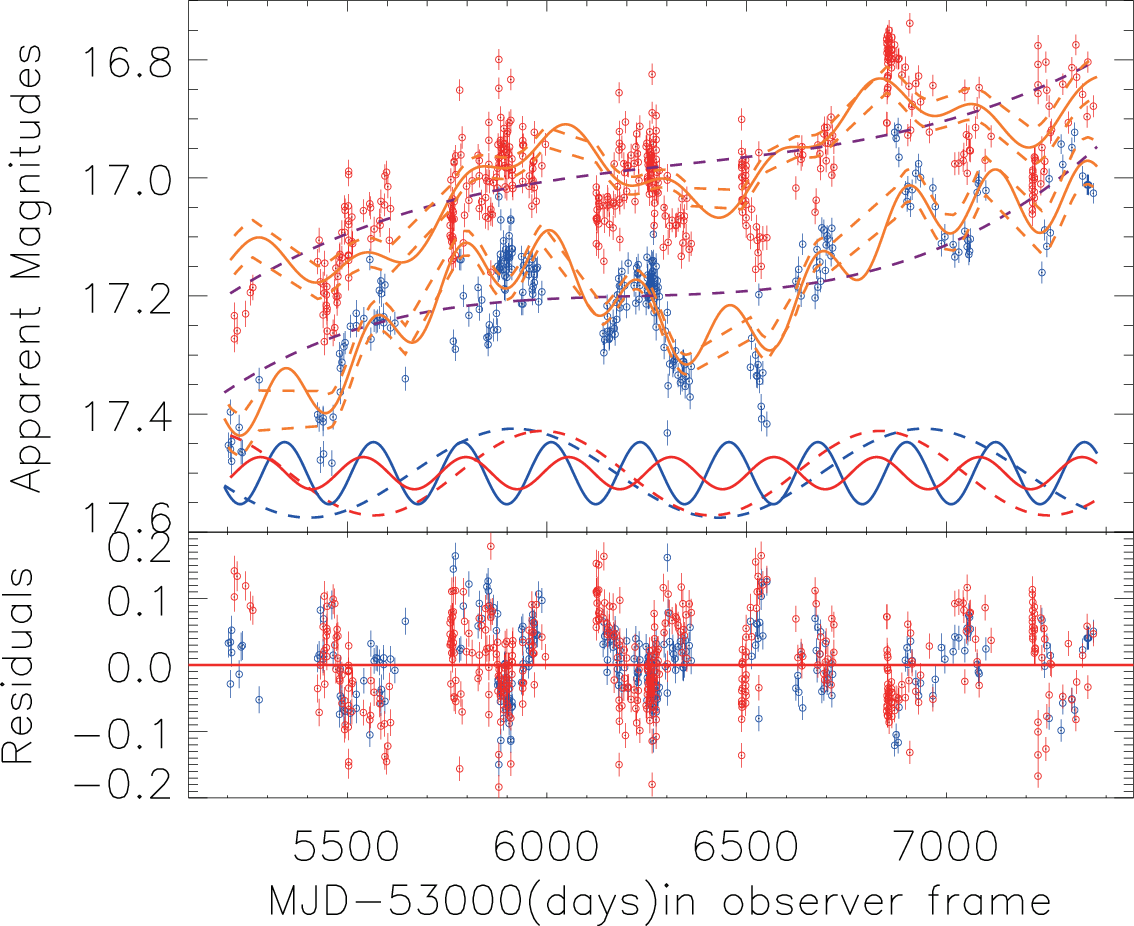}
\caption{Best fitting results (top panel) and the corresponding residuals (bottom panel) to the ZTF g/r-band light curves by two 
sinusoidal components. In top panel, circles plus error bars in blue and in red show the ZTF g-band and r-band light curves, the 
solid orange lines represent the best-fitting results, the dashed orange lines indicate the 5$\sigma$ (99.99994\%) confidence 
bands, and the  dashed purple lines show the polynomial components, the dashed blue and solid blue lines show the sinusoidal 
components with periods of $1024\pm12$days and $222\pm1$days in the ZTF g-band light curve, the dashed and solid lines in red
represent the sinusoidal components with periods of $886\pm6$days and $257\pm1$days in the ZTF r-band light curve. }
\label{ztfg}
\end{figure}

	Based on the secondary periodicity determined by the GLS method and the WWZ method, especially in the ZTF light curves, 
Figure~\ref{ztfg} shows the re-fitting results to the ZTF g-band and r-band light curves by using two sinusoidal functions. The model 
functions are expressed as follows:
\begin{equation}
\begin{split}
LMC{2} &= A + B \times \left(\frac{t}{1000 \text{ days}}\right) +\quad C \times \left(\frac{t}{1000 \text{ days}}\right)^2\\
	&\quad + D \times \left(\frac{t}{1000 \text{ days}}\right)^3 
	\quad+ E \times \sin\left(\frac{2\pi t}{T_{QPO}} + \phi_0\right)\\
	&\quad + F \times \sin\left(\frac{2\pi t}{T_{QPO_1}} + \phi_1\right)	
\end{split}
\end{equation}
Then, through the Levenberg-Marquardt least-squares minimization technique, the best fitting results can be determined and shown in 
Figure~\ref{ztfg} to the ZTF light curves, leading to $\chi^2/\text{dof} = 7430.6168/728 \approx 10.206$. The determined model 
parameters listed in Table~\ref{table1} can be applied to support the secondary periodicities in the ZTF g/r-band light curves. 
Based on the model function used to fit the light curves in the left panel of Figure \ref{mcmc}, when only the ZTF g/r bands are 
fitted, the resulting $\chi^2/\text{dof}$ is calculated as 8767.5/734 $\approx 11.945$. As demonstrated in Equation (2) and (3) 
above, the F-test technique is employed to compare the model containing one sinusoidal component with the model containing two 
sinusoidal components. The result of $F_{P}$ is about $F_{P}\approx21.8 $, and the $F_{dis}$ is about $F_{dis}\approx 5.8 $ (based 
on the given confidence level($5\sigma$)). These results indicate that the model with two sinusoidal components provides a better fit.

	By applications of four different techniques, including the direct fitting method leading to the results shown in the left 
panel of Figure \ref{mcmc} and Figure \ref{ztfg}, the phase-folded method leading to the results shown in the right panel of 
Figure \ref{mcmc}, the GLS periodogram leading to the results shown in the Figure \ref{gls}, and the WWZ method leading to the 
results shown in the Figure \ref{wwz}. All these methods are leading to two significant periodicities in ZTF light curves. Meanwhile, 
for the CSS-V band light curve, a significant periodicity is detected by four methods, while an insignificant periodicity is 
detected by The WWZ method and GLS. Differences in detected short-period signals between the CSS and ZTF bands may be attributed 
to the higher temporal resolution and observational precision of ZTF light curves, which can reveal finer variability features that 
the CSS V-band might have missed. In summary, there are two periodicities in \obj, at $1103\pm260$days (1103 as the mean value of 
the longer periodicities of 985days, 1024days, 886days, 1250days, 1165days, 1000days, 1265days, 1405days and 950days from 
different techniques applied to different light curves, and the uncertainty 260 as half of the maximum difference among the longer 
periodicities) days and $243\pm29$days (243 as the mean value of the shorter periodicities of 222days, 257days, 272days, 221days, 
215days, 265days, 225days, 270days, and the uncertainty 29 as half of the maximum difference among the shorter periodicities). 
The source of the above periodicities is shown in table\ref{table3}.

	\subsection{Optical QPOs were related to intrinsic AGN variability of \obj?}

	Similar to what we have recently done in \citet{zh23b, zh25, zh25b}, it is necessary to check the effects of red noise traced by 
intrinsic AGN activities on the detected optical QPOs in \obj, accepted artificial light curves $LMC_t$ (red noises, intrinsic AGN variability) 
being simulated by the Continuous Autoregressive (CAR) process \citep{kbs09, m10, k10, z13}:
\begin{equation}
dLMC_t = \frac{-1}{\tau} LMC_t \, dt + \sigma_c \sqrt{dt} \epsilon(t) + bdt
\end{equation}
where $\epsilon(t)$ is a white noise process with zero mean and variance equal to 1, $bdt=17.19$ is the mean value of $LMC_t$ (with 17.19 as 
the mean value of the ZTF g-band light curve of \obj), $\tau$ is the relaxation time of the process, and $\sigma$ is the variability of the 
time series on a timescale shorter than $\tau$. And the uncertainties of the simulated light curves $LMC_{t}$ are given by 
$dLMC_{t,err} = LMC_t \ + \frac{L_{err}} {L_{obs}}$, with $L_{obs}$ and $L_{err}$ as the observed ZTF g-band light curve and the corresponding 
uncertainties, as shown in the left panel of Figure \ref{mcmc}.

	Through the CAR process in \citet{kbs09}, two sets of 10000 simulated light curves are generated. The first set had the same time 
information $t$ as the actual observation time of CSS V band and ZTF g band light curves, with $\tau$ randomly selected from 50days to 
1000days and the variance $\tau$$\sigma_*^2$ from 0.005 to 0.02 as the common values in quasars in \citet{m10}. The second set had the 
same time information as the observation time of ZTF g band light curve, with $\tau$ from 50days to 1000days and the variance 
$\tau$$\sigma_*^2$ from 0.005 to 0.02.

\begin{figure}
\centering\includegraphics[width = 8cm,height=6cm]{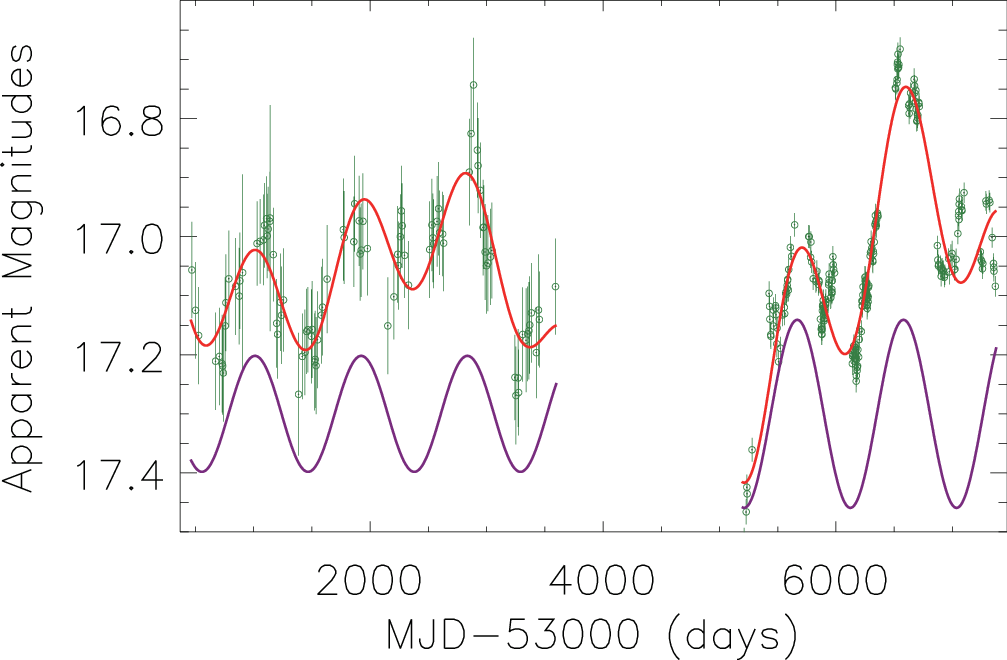}
\centering\includegraphics[width = 8cm,height=6cm]{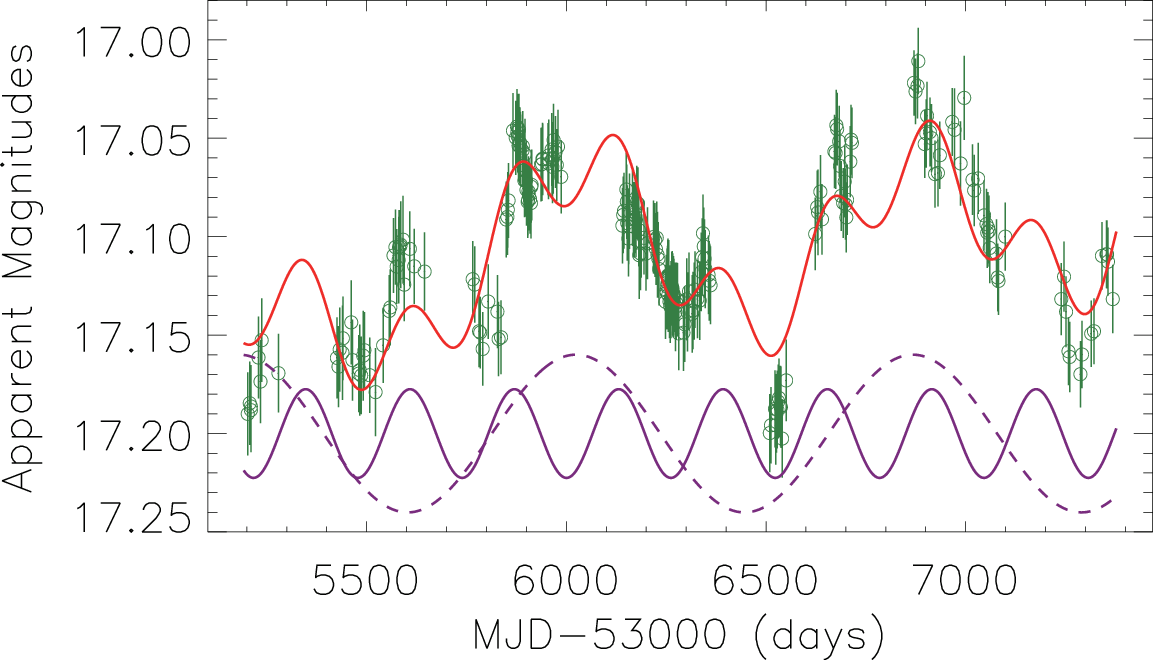}
\caption{The top panel shows an example of probable mis-detected QPOs in the first set of simulated light curves by the CAR 
process. Solid dark green circles plus error bars show the simulated light curve, solid lines in red show the best-fitting result by Equation(1), 
with $\chi^2/\text{dof}$=6. The solid lines in purple show the sinusoidal component with a period of 910days. The bottom panel shows one 
of the probable mis-detected QPOs in the second set of simulated light curves. Solid dark green circles plus error bars show the simulated 
light curve, solid line in red shows the best-fitting result by Equation(5), with $\chi^2/\text{dof}$=1.13. The solid and dashed lines in 
purple show the sinusoidal components with periodicity of 263days and 851days, respectively.}
\label{drw}
\end{figure}

	There are 15 light curves with detected QPOs that can be selected from the first set of simulated light curves, based on the following 
two criteria. First, the simulated light curves can be well described by Equation (1) with corresponding $\chi^2/\text{dof}$ smaller than 15 
($\chi^2/\text{dof}$=10.69 in Fig \ref{mcmc}), and the determined periodicity is within the range of 1103$\pm$260days. Second, the periodicity 
determined by the GLS method should be within 1103$\pm$260days with  significance level higher than 5$\sigma$. For the second set of simulated 
light curves, 2 light curves with expected detected QPOs can be selected if they meet the following two conditions. First, the simulated light 
curves must be well described by Equation (5) with corresponding $\chi^2/\text{dof}$ smaller than 15 ($\chi^2/\text{dof}$=10.206 in Fig \ref{ztfg}), 
and the determined two periodicities are within 1103$\pm$260days and 243$\pm$29days. Second, the periodicities determined by the GLS method 
should be within 1103$\pm$260days and 243$\pm$29days, with significance level higher than 5$\sigma$. One example of these mis-detected QPOs 
from each set of simulated light curves treated as red noises is shown in Figure \ref{drw}.

	The results above indicate that the CAR process, which can be applied to trace red noise, can lead to mis-detected QPOs in simulated light 
curves. However, after considering effects of red noise, it can be confirmed that the probability is only 2/10000 (corresponding $3.7\sigma$) 
for two periodicities detected over the ZTF observation time, and the probability is only 15/10000 (corresponding 3.17$\sigma$) for only the large 
periodicity over the combined CSS and ZTF observation time. Therefore, the detected QPOs in \obj~ are robust enough, even after considering effects 
of red noise.

\begin{table}
\caption{Parameter values for the LMC model function}
\begin{tabular}{llll}
\hline\hline
			LMC$_{\text{1}}$ & CSS-V & g band & r band \\
			\hline
			A & $16.7 \pm 0.09$ & $56.2 \pm 1.74$ & $21.7 \pm 1.58$ \\
			B & $-0.02 \pm 0.17$ & $-18.7 \pm 0.84$ & $-1.97 \pm 0.76$ \\
			C & $-0.02 \pm 0.09$ & $3.02 \pm 0.13$ & $0.27 \pm 0.12$ \\
			D & $0.01 \pm 0.01$ & $-0.16 \pm 0.01$ & $-0.01 \pm 0.01$ \\
			E & $-0.06 \pm 0.01$ & $0.06 \pm 0.01$ & $0.06 \pm 0.01$ \\
			$\log T_{QPO}$ & $2.99 \pm 0.01$ & $2.99 \pm 0.01$ & $3 \pm 0.01$ \\
			\(\phi_0\) & $-0.39 \pm 0.22$ & $-1.56 \pm 0.31$ & $-1.44 \pm 0.31$ \\
			\hline
			
			LMC$_{\text{2}}$ & CSS-V & g band & r band \\
			\hline
			A & & $51.3 \pm 2.52$ &  $40.6 \pm 1.79$ \\
			B & & $-16.5 \pm 1.21$& $-11.04 \pm 0.85$  \\
			C & & $2.65 \pm 0.20$ &  $1.73 \pm 0.14$  \\
			D & & $0.14 \pm 0.01$ &  $-0.09 \pm 0.01$ \\
			E & & $0.08 \pm 0.01$ &  $ -0.07 \pm 0.01$ \\
			$\log(T_{QPO})$ & & $3.01 \pm 0.01$ & $2.9 \pm 0.01$  \\
			\(\phi_0\) & & $6.02 \pm 0.48$ & $ -5.25 \pm 0.32$ \\
			F & & $0.05 \pm 0.01$ & $0.03 \pm 0.01$  \\
			$\log(T_{QPO_1})$ & & $2.35 \pm 0.01$ & $ 2.41 \pm 0.01$ \\
			\(\phi_1\) & & $-20.5 \pm 0.47$ & $1.3 \pm 0.78$ \\ 
			\hline
			LMC$_{\text{ph}}$  & A & B & C \\
			\hline
			& $-0.06\pm 0.01$ & $-0.60\pm0.02$ & $-0.01\pm0.01$\\
			\hline
		\end{tabular}\\
		\tablefoot{The first part shows the determined parameters after applications of the model functions of LMC$_{\text{1}}$ 
			in Equation (1) to the CSS and ZTF light curves. The second and the third part show the determined parameters after 
			applications of the model functions of LMC$_{\text{2}}$ in Equation (5) and of LMC$_{\text{ph}}$ in Equation (4) to 
			the CSS and ZTF light curves. \\
			The parameters of $T_{QPO}$ and $T_{QPO_1}$ are the periodicities in units of days.
		}
		\label{table1}
	\end{table}

\section{spectroscopic results of \obj}

\begin{figure}
\centering\includegraphics[width = 8cm,height=5.5cm]{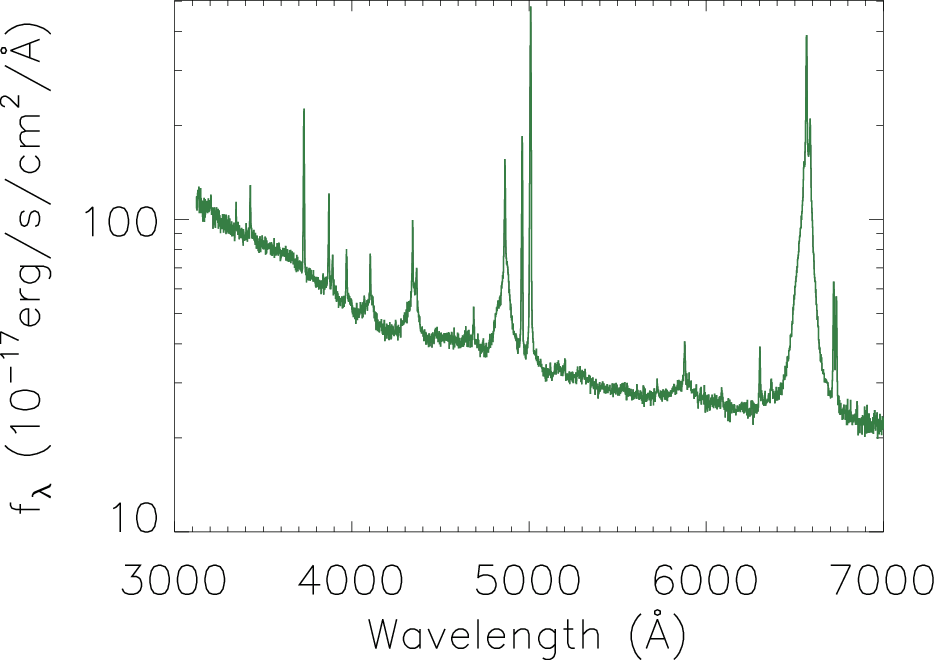}
\caption{The SDSS spectrum of \obj.
}
\label{view}
\end{figure}

	Considering that the optical QPOs may be caused by orbital motions of two BH accreting systems in an expected central 
BSMBH system in \obj, this orbital motions could also have probable effects on features of broad optical Balmer emission 
lines, if the related two broad emission line regions (BLRs) were not totally mixed. Therefore, it is necessary to study the 
characteristics of the broad emission lines. Figure \ref{view} displays the SDSS spectrum of \obj, collected from SDSS DR16 \citep{ah21} 
with PLATE-MJD-FIBERID = 2586-54169-0405. In this spectrum, prominent emission lines can be described by multiple Gaussian functions, 
as demonstrated by \citet{zh22a, zh23b}, which utilized a combination of broad and narrow Gaussian functions to precisely measure the 
emission lines.
	
	\begin{table}
		\centering
		\caption{The periodicities original}
		\begin{tabular}{llll}
			\hline                                                                                      
			method & CSS-V & ZTF-g &  ZTF-r \\
			\hline
			direct fit & $985\pm8$ & * & * \\
			\hline
			direct re-fit &  & $1024\pm12$ & $886\pm6$ \\
			& & $222\pm1$ & $257\pm1$ \\		
			\hline
			GLS & $1250\pm112$ & $1165\pm30$ & $1000\pm30$ \\
			& $272\pm1$ & $221\pm1.5$ & $215\pm1$ \\
			\hline
			WWZ& $1265\pm40$ & $1405\pm40$ & $950\pm42$ \\
			& $265\pm3$ & $225\pm3$ & $270\pm3.5$ \\
			\hline

		\end{tabular}\\
		\tablefoot{Repeated periodicities are not used for calculation again.}
		\label{table3}
	\end{table}
	
\begin{figure*}
\centering\includegraphics[width = 8cm,height=5.5cm]{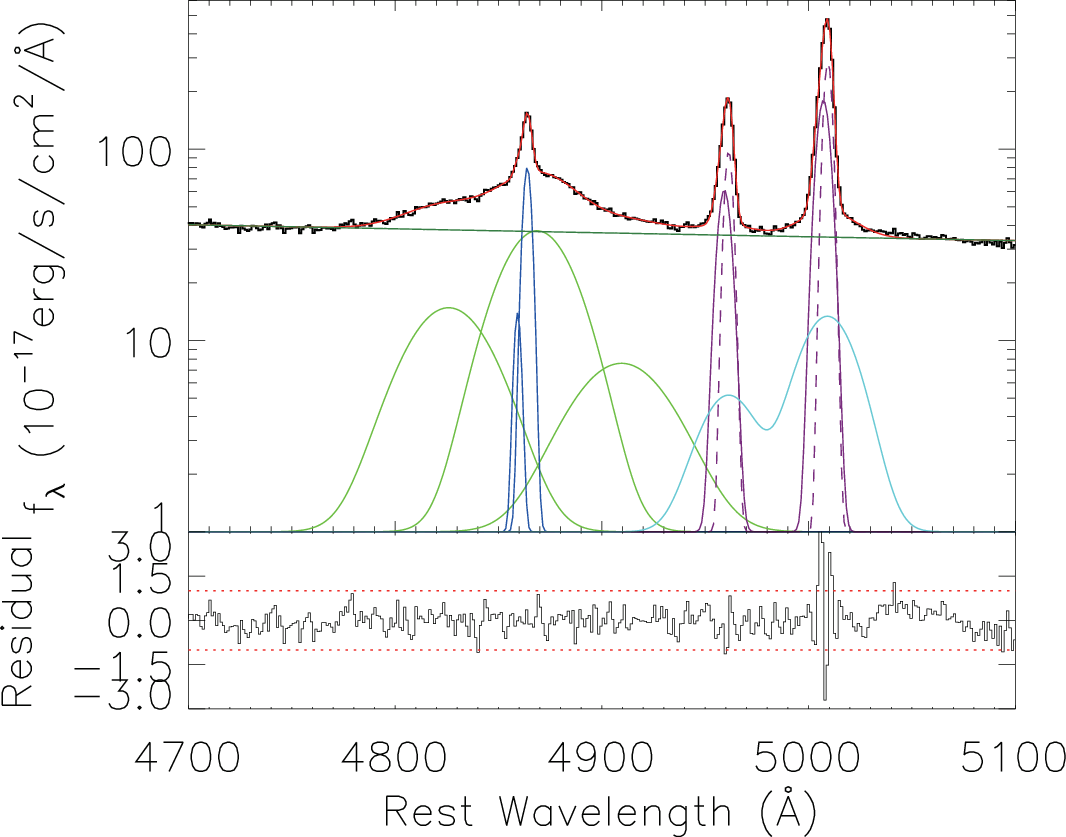}
\centering\includegraphics[width = 8cm,height=5.5cm]{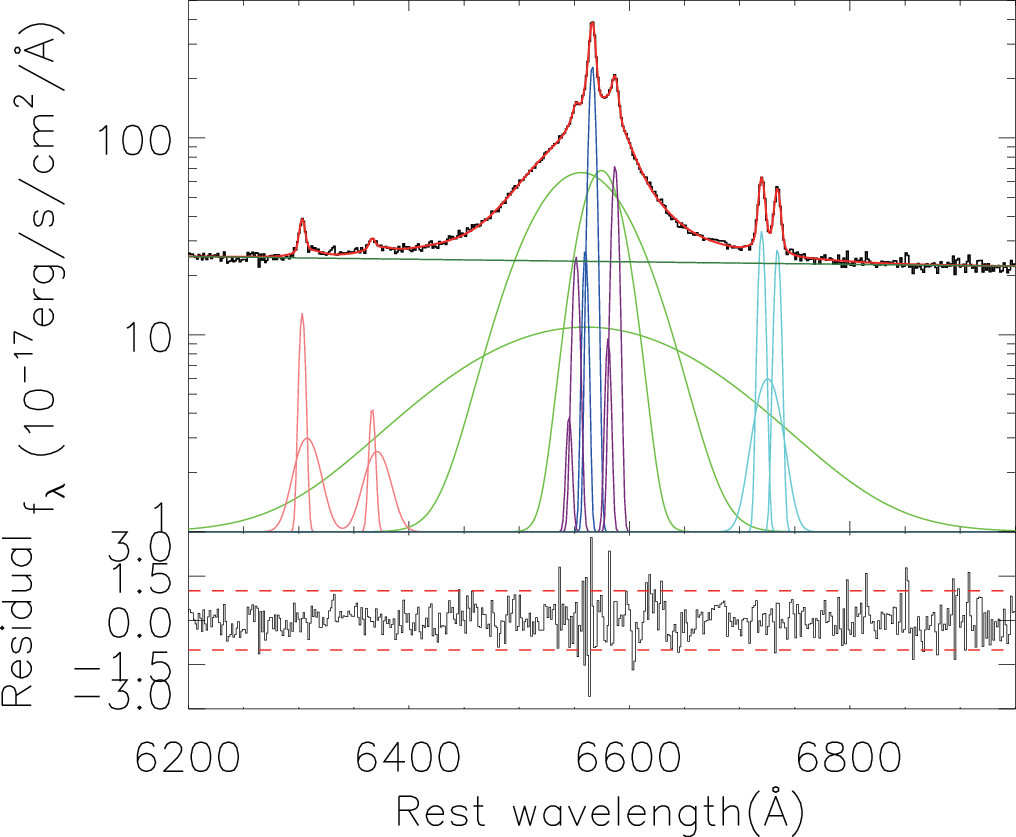}
\centering\includegraphics[width = 8cm,height=6cm]{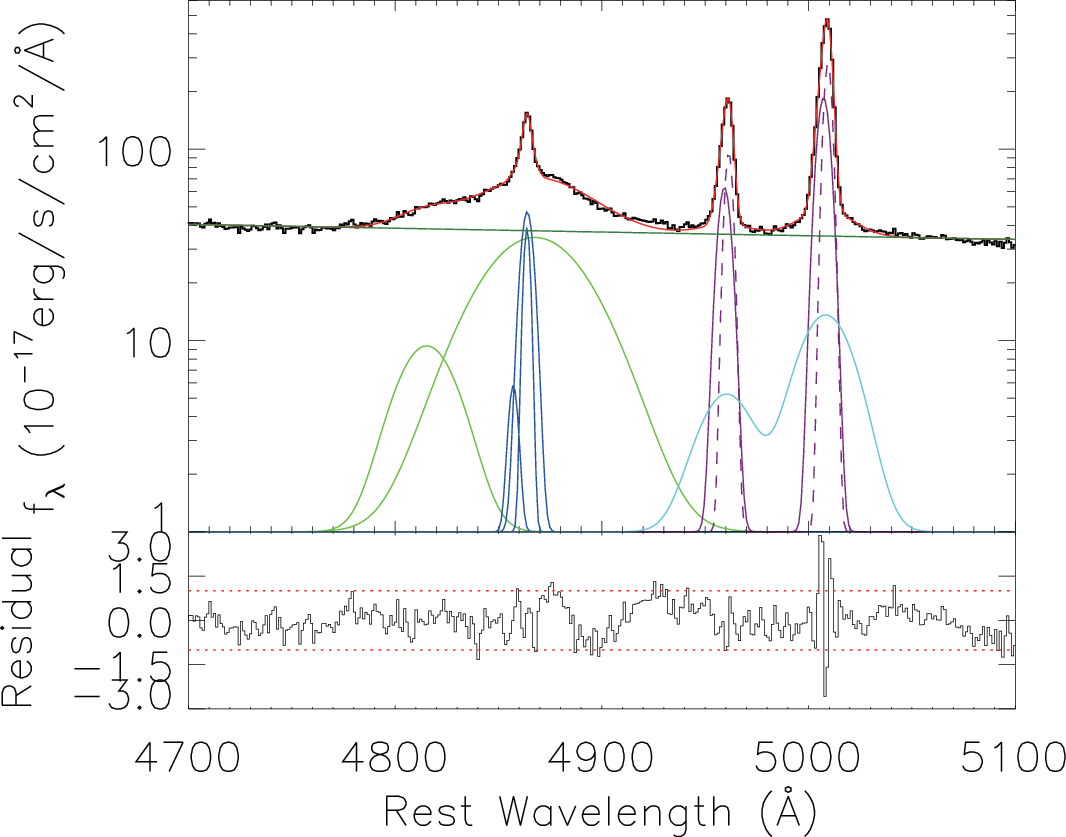}
\centering\includegraphics[width = 8cm,height=6cm]{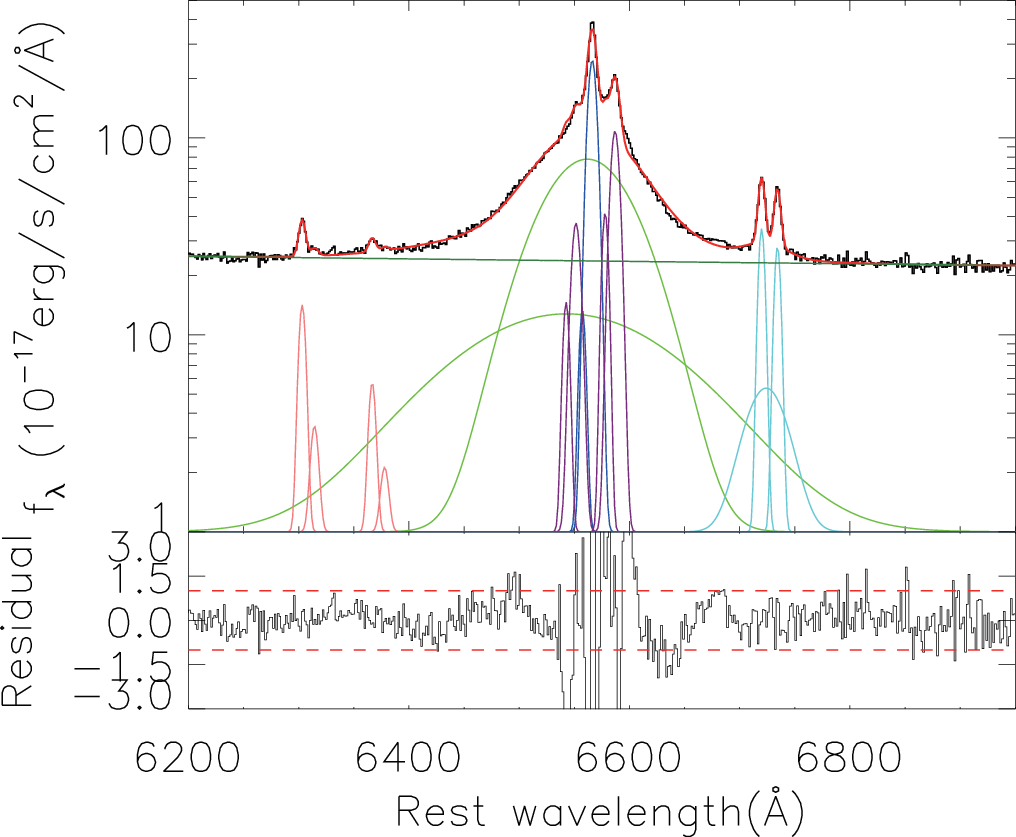}
\caption{The top panels show the fitting results and the corresponding residuals for the emission lines around $H\beta$ and 
$H\alpha$, respectively, considering three broad Gaussian components for broad Balmer lines. In the top left panel, the solid black line 
represents the observed SDSS spectrum, the solid red line indicates the best fitting results, the green and blue lines show the broad and 
narrow H$\beta$ components, the solid cyan and purple lines represent the core and extended components of the [O~{\sc iii}] doublet. 
In the top right panel, the solid black line represents the observed SDSS spectrum, the solid red line represents the best fitting results, 
the solid green line illustrates the broad Gaussian components of H$\alpha$, the solid blue line represents the narrow Gaussian components 
in H$\alpha$, the solid purple line shows the core and extended components of the [N~{\sc ii}] doublet, the solid cyan line represents 
the core and extended components in the [S~{\sc ii}] doublet, the solid pink line indicates the core and extended components in the [O~{\c i}] 
doublet. In each top panel, the dashed green line represents the determined power-law continuum emissions. For each residuals, the 
horizontal dashed red lines show residuals=$\pm1$. In order to show clear features of the determined emission components, the plots are 
shown with y-axis in logarithmic. Similar to the results presented in top panels, bottom panels show the fitting results 
and the corresponding residuals for the emission lines around H$\beta$ and H$\alpha$, respectively, considering two broad Gaussian components 
for broad Balmer lines.}
\label{fit}
\end{figure*}

	To measure the emission lines around H$\beta$ (rest wavelength from 4700\AA\ to 5100\AA), the following model functions are 
utilized. For the broad and narrow H$\beta$ emission lines, three broad Gaussian functions and two narrow Gaussian functions (one for the 
core component and the other one for the extended component) are applied. For the [O~{\sc iii}] $\lambda$4959\AA, 5007\AA \ doublet, three 
Gaussian functions are applied to each line, to describe the core component, the extended component and an additional intermediate 
extended component. Additionally, a power-law function is applied to describe the continuum emissions underneath the emission lines. 
Then, through the Levenberg-Marquardt least-squares minimization technique, the best fitting results and the corresponding residuals to 
the emission lines around H$\beta$ are determined and shown in the top left panel of Figure~\ref{fit} with $\chi^2/\text{dof} = 1.1$. 
Here, the corresponding residuals are determined by the spectrum minus the best fitting results, and then divided by the uncertainties 
of the spectrum. The determined model parameters are listed in Table~\ref{table2}.

	\begin{table}
		\centering
		\caption{Line parameters}
		\begin{tabular}{llll}
			\hline
			line & $\lambda_0$ & $\sigma$ &  flux \\
			\hline
			Broad H$\alpha$ & $6556.3\pm0.5$ & $41.6\pm0.6$ & $6842\pm167$ \\
			& $6560.5\pm2.4$ & $106.8\pm4.9$ & $2663\pm125$ \\
			& $6574.7\pm0.4$ & $17.5\pm0.4$ & $2956\pm127$ \\
			\hline
			Broad H$\beta$ & $4868.7\pm1.1$ & $16.9\pm1.2$ & $1549\pm207$ \\
			& $4825.8\pm3.2$ & $19.6\pm1.9$ & $679\pm103$ \\
			& $4909.5\pm8.8$ & $22.1\pm4.8$ & $365\pm140$ \\
			\hline
			Narrow H$\alpha$ & $6566.3\pm0.1$ & $2.7\pm0.1$ & $1592\pm55$ \\
			& $6559.6\pm0.6$ & $2.3\pm0.4$ & $149\pm44$ \\
			\hline
			Narrow H$\beta$ & $4863.9\pm0.1$ & $2.0\pm0.1$ & $403\pm15$ \\
			& $4859.2\pm0.4$ & $1.7\pm0.3$ & $54\pm13$ \\
			\hline
			\multirow{2}{*}{[O~{\sc iii}]$\lambda5007$\AA}& $5009.3\pm0.1$ & $2.0\pm0.1$ & $1442\pm122$ \\
			& $5007.2\pm0.3$ & $3.1\pm0.1$ & $1406\pm126$ \\
			& $5009.2\pm0.5$ & $12.6\pm0.6$ & $402\pm18$ \\
			\hline
			[O~{\sc i}]$\lambda6300$\AA & $6303.2\pm0.2$ & $2.5\pm0.2$ & $74\pm7$ \\
			& $6307.5\pm1.8$ & $10.6\pm1.9$ & $53\pm12$ \\
			\hline
			[O~{\sc i}]$\lambda6363$\AA & $6366.6\pm0.2$ & $2.5\pm0.2$ & $21\pm5$ \\
			& $6371.1\pm1.8$ & $10.7\pm1.9$ & $42\pm12$ \\
			\hline
			[S~{\sc ii}]$\lambda6716$\AA & $6719.9\pm0.1$ & $2.7\pm0.1$ & $222\pm13$ \\
			& $6725.4\pm1.3$ & $10.9\pm1.1$ & $137\pm26$ \\
			\hline
			[S~{\sc ii}]$\lambda6731$\AA & $6734.3\pm0.1$ & $2.7\pm0.1$ & $180\pm12$ \\
			& $6739.8\pm1.3$ & $10.9\pm1.1$ & $0*$ \\
			\hline
			[N ~{\sc ii}]$\lambda6583$\AA & $6586.9\pm0.1$ & $2.7\pm0.1$ & $491\pm23$ \\
			& $6580.6\pm0.6$ & $2.3\pm0.4$ & $49\pm19$ \\
			\hline
			\hline
			line & $\lambda_0$ & $\sigma$ &  flux \\
			\hline
			Broad H$\alpha$ & $6562\pm0.3$ & $39.4\pm0.5$ & $7610\pm140$ \\	
			& $6543\pm2.3$ & $93.1\pm3.3$ & $2750\pm145$ \\
			
			\hline
			Broad H$\beta$ & $4867.6\pm0.5$ & $25.0\pm0.4$ & $2108\pm39$ \\
			& $4815.3\pm1.2$ & $13.7\pm1.1$ & $288\pm32$ \\
			
		   \hline
	    	Narrow H$\alpha$ & $6566\pm0.1$ & $3.9\pm0.1$ & $2436\pm36$ \\
	    	& $65560\pm0.5$ & $3.9\pm0.4$ & $0*$ \\
	    	& $6557\pm0.6$ & $2.8\pm0.6$ & $86\pm23$ \\
			\hline
			Narrow H$\beta$ & $4863.7\pm0.1$ & $1.6\pm0.1$ & $154\pm16$ \\
			& $4857.1\pm0.2$ & $2.2\pm0.1$ & $27\pm7$\\
			& $4863.8\pm0.1$ & $3.0\pm0.1$ & $346\pm21$\\
	    	\hline
	    	
	    	\multirow{2}{*}{[O~{\sc iii}]$\lambda5007$\AA} & $5009.3\pm0.1$ & $2\pm0.2$ & $1402\pm122$ \\
	    	& $5007.2\pm0.2$ & $3.1\pm0.1$ & $1438\pm126$ \\
	    	& $5008.3\pm0.4$ & $12.5\pm0.6$ & $396\pm18$ \\
	    	\hline
	    		[O~{\sc i}]$\lambda6300$\AA & $6303.1\pm0.2$ & $3\pm0.2$ & $98\pm5$ \\
	    	& $6314.3\pm0.9$ & $3.4\pm1$ & $21\pm5$ \\
	    	\hline
	    	[O~{\sc i}]$\lambda6363$\AA & $6366.6\pm0.2$ & $3.0\pm0.2$ & $35\pm4$ \\
	    	& $6377.9\pm1.8$ & $13.4\pm1$ & $10\pm4$ \\
	    	\hline
	    	[S~{\sc ii}]$\lambda6716$\AA & $6719.9\pm0.1$ & $2.9\pm0.1$ & $239\pm9$ \\
	    	& $6724\pm1.6$ & $17.9\pm1.1$ & $196\pm24$ \\
	    	\hline
	    	[S~{\sc ii}]$\lambda6731$\AA & $6734.3\pm0.1$ & $2.8\pm0.1$ & $192\pm8$ \\
	    	& $6738.4\pm1.3$ & $18\pm1.1$ & $0*$ \\
	    	\hline
	    	[N ~{\sc ii}]$\lambda6583$\AA & $6586.9\pm0.1$ & $4\pm0.1$ & $1063\pm26$ \\
	    	& $6577.5\pm0.5$ & $2.8\pm0.4$ & $284\pm25$ \\
	    	\hline	
		\end{tabular}\\
		\tablefoot{The first column shows which line is measured. The second, third, fourth column show the measured 
		line parameters: the center wavelength $\lambda_0$ in units of \AA, the line width (second moment) $\sigma$ in units 
		of \AA~ and the line flux in units of ${\rm 10^{-17}~erg/s/cm^2}$. \\	
		The first part presents the parameters when the broad Balmer lines are fitted with three broad Gaussians, 
		and the second part presents the parameters when fitted with two broad Gaussians. \\
		For [S~{\sc ii}]$\lambda6731$\AA, there are two Gaussian components applied, but only one reliable Gaussian component 
		is determined and shown in the right panels of Figure~\ref{fit}\\
		For the second part of the narrow H$\alpha$, there are three Gaussian components applied, but only two 
		reliable Gaussian components are determined and shown in the bottom right panel of Figure~\ref{fit}} 
		
		\label{table2}
	\end{table}

	For the emission lines around H$\alpha$ (rest wavelength from 6200\AA\ to 6900\AA), The following model functions are applied. 
Five Gaussian functions are applied to describe the broad and narrow H$\alpha$ emission lines, with three Gaussian functions for the 
broad H$\alpha$ and two for the narrow H$\alpha$. The [N~{\sc ii}] $\lambda$6548\AA, 6583\AA\ doublet are described by four Gaussian functions 
for both the core and extended components. The [S~{\sc ii}] $\lambda$6716\AA, 6731\AA\ doublet and the [O~{\sc i}] $\lambda$6300\AA, 6363 \AA\ 
doublet are also fitted in a similar manner, with one Gaussian function for core component and the other one Gaussian function for 
the extended component in each line. A power-law function is used for the continuum emissions underneath the emission lines. Then 
through the Levenberg-Marquardt least-squares minimization technique, the best fitting results and the corresponding residuals are 
determined and shown in the top right panel of Figure~\ref{fit}, with $\chi^2/\text{dof} = 1.1$. The determined model parameters are 
listed in Table~\ref{table2}.

\begin{figure}
\centering\includegraphics[width =\columnwidth]{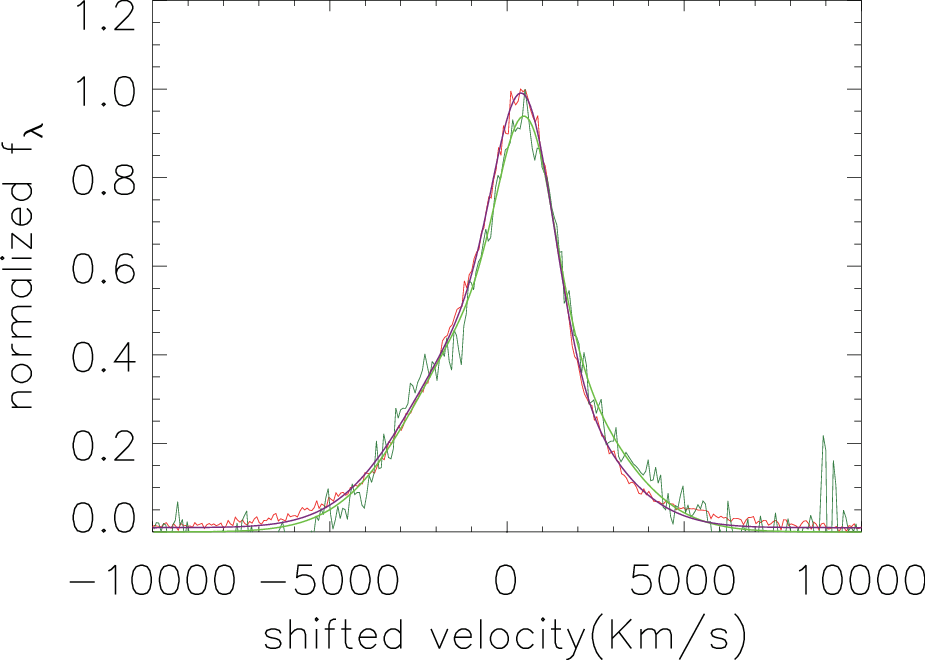}
\caption{Line profile Comparisons of broad H$\alpha$ (in red) and broad H$\beta$ (in green) in velocity space, after subtracting the 
continuum and all other narrow emission lines.
	}
\label{fake}
\end{figure}

	To further analyze the significant physical information in the broad emission lines, the continuum and the narrow emission lines 
are subtracted from the spectra, leaving only the broad emission components of H$\alpha$ and H$\beta$. Then, the normalized broad H$\alpha$ 
and H$\beta$ are shown in velocity space in Figure~\ref{fake}, to show line profile comparisons between broad H$\alpha$ and broad 
H$\beta$ in \obj. Meanwhile, based on the second moment definition of emission lines \citep{pe04}, the measured line widths of the broad 
H$\alpha$ and broad H$\beta$ are $2677.8\pm539$ km/s and $1871.67\pm441.24$ km/s (the uncertainty determined by the maximum error of 
$\sigma$ for the broad Gaussian components in Table \ref{table2}), respectively. It is evident that the line width of broad H$\alpha$ 
is a bit greater than that of broad H$\beta$. However, as noted by \citet{kg04, bk10, nh20}, broad H$\alpha$, having the largest intrinsic 
optical depth among the broad Balmer emission lines, would typically exhibit a narrower (or similar, but not larger) line width than of 
broad H$\beta$. This suggests that intrinsic optical depth differences alone cannot explain the observed line profile variations. As 
proposed by \citet{zh23a}, the distinct line profiles of broad Balmer emission lines are likely due to extrinsic obscuration differences 
between the two independent broad-line regions of the BSMBH system, providing support for the presence of central 
BSMBHs. Additionally, the peak velocities (relative to the rest-frame central wavelength) of H$\alpha$ and H$\beta$ are 
approximately $-85.59\pm110.59$ km/s and $241.18\pm543$ km/s, respectively (the uncertainty determined by the maximum error of 
$\lambda_0$ for the broad Gaussian components in Table \ref{table2}). The significant shifted velocity difference also implies that 
if the broad H$\alpha$ and broad H$\beta$ originated from the probably different regions otherwise similar kinematic system should lead 
to similar peak velocities and similar line widths (second moment). Therefore, the results above suggest the possible existence of two 
broad emission line regions.

	Furthermore, the intensity ratio of the broad H$\alpha$ to broad H$\beta$ can be calculated as 4.83, a bit larger than the 
theoretical value of 2.86, indicating that dust extinction might be present in the BLRs of \obj, causing more absorption of H$\beta$. 
However, while dust extinction could explain the larger intensity ratio than 2.86, it alone is insufficient to account for the velocity 
and profile differences between the broad H$\alpha$ and broad H$\beta$. Moreover, differences in optical depth may partially contribute 
to the differences in the profiles of broad H$\alpha$ and broad H$\beta$, but they cannot solely explain the significant velocity 
shifts between broad H$\alpha$ and broad H$\beta$.

	In conclusion, the hypothesis of a BSMBH system is a more reasonable explanation: broad H$\alpha$ and broad H$\beta$ 
originate from two distinct BLRs. Future work will require long-term spectral monitoring to search for periodic red-shifted and 
blue-shifted features, multi-wavelength observations to analyze dust distribution, and kinematic modeling to further confirm the 
existence of this BSMBH system.

\section{discussion}

\subsection{Discussions on the expected BSMBH system}
	
	If there was an expected BSMBH system in \obj, the discussion of BH mass becomes crucial. Typically, the virial mass 
of the BH can be estimated from the properties of two broad components in the broad Balmer lines. In this case, the H$\beta$ and H$\alpha$ 
lines are refitted with two prominent Gaussian components and three narrow Gaussian components, respectively. The fitting results are 
shown in bottom panels of Figure \ref{fit}, and the determined parameters are listed in Table \ref{table2}. Here, we should note that 
the main purpose of the refitted results is only to check dynamical properties under the assumption that there were two independent BLRs 
related to each BH accretion system in the expected BSMBH system in \obj. Moreover, as shown in bottom panels of Figure \ref{fit} and the 
determined parameters listed in Table \ref{table2}, one of the two determined broad components in broad H$\alpha$ has no shifted velocities, 
therefore, the determined parameters of the two broad Gaussian components in broad H$\beta$ are mainly considered as follows.

	If the refitted two broad components do originate from two independent BLRs, the masses of the two BHs in the central BSMBH system 
can be estimated as follows. According to the virial theorem, the kinetic energy of the gas cloud balances with its gravitational potential 
energy \citep{pe04, vp06}. Combining this with the empirical R-L relation \citep{wp99, ks00, bd13}, which shows a strong linear correlation 
between the size of the BLRs and the optical continuum luminosity (or the broad line luminosity), the virial BH mass can be expressed as 
\citep{pe04, gh05}:
\begin{equation}
	\frac{M_{BH}}{M_{\odot}} = 2.2 \times 10^{6} \times \left( \frac{L_{H\beta}}{10^{42} \text{ erg/s}} \right)^{0.56} \times \left( \frac{FWHM_{H\beta}}{1000 \text{ km/s}} \right)^{2.06}
\end{equation}
with the line widths and the line luminosities of the two broad Gaussian components of the H$\beta$, the masses of the redshifted and 
blueshifted BH systems are determined to be $m_{\rm BHr}$=$(5.8\pm0.28)\times10^{7} M_\odot$ and $m_{\rm BHb}$=$(5.49\pm1.37)\times10^{6} M_\odot$, 
respectively. 

	The mass of redshifted BH system is far greater than that of the blueshifted BH system. In this case, the orbital separation 
between the two BHs can be simply estimated according to Kepler's laws, with the blueshift velocity and the mass of the primary SMBH:
\begin{equation}
	r = \frac{GM_{\rm BHr}}{(V_{b})^2}\approx(9.55\pm1.2) \times 10^{14}\text{m} = (1.12\pm0.14)\times10^4 R_G
\end{equation}
Subsequently, the orbital period can be calculate as:
\begin{equation}
	T = \sqrt{\frac{4\pi^{2}r^{3}}{GM_{\rm BHr}}}\approx (2.1\pm0.4) \times 10^{9}\text{s}=66.9\pm1.2 yr
\end{equation}
The orbital period is approximately 66.9 years, showing a certain discrepancy with the detected optical QPOs in \obj. However, this discrepancy may stem from the unreliable re-fitting of the two prominent Gaussian components for the broad Balmer lines, especially when compared to the three components model. The $\chi^2/\text{dof}$ value of the two components model is higher than that of the three components model (1.8 $>$  1.1), and the residuals of the two components model are larger and more scattered.  In addition, the discrepancy may also result from the mixing of BLRs, where the two broad components may not originate from two independent BLRs. As a result, the calculated results may not be fully reliable based on these two broad components. therefore, the virial mass of the BH estimated from the properties of two broad components is not used in this manuscript.

	Instead, the BH mass is determined based on the broad Balmer lines using Equation 7 considering the line parameters determined by 
three Gaussian functions to describe the broad Balmer lines, with the line luminosity of total broad H$\alpha$ 
$L_{H\alpha}\sim(1.76\pm0.05)\times10^{43}{\rm erg/s}$ and of total broad H$\beta$ $L_{H\beta}\sim(0.37\pm0.06)\times10^{43}{\rm erg/s}$ 
and FWHM of total broad H$\alpha$ $FWHM_{H\alpha}\sim3102\pm155$km/s and of total broad H$\beta$ $FWHM_{H\beta}\sim2936\pm150$km/s), the 
virial BH masses can be calculated as \( M(\text{BH,H}\alpha)\sim(1.13\pm0.14)\times10^8 M_\odot \) and 
\( M(\text{BH,H}\beta)\sim(4.17\pm0.9)\times 10^7 M_\odot \). There is an obvious difference in the BH mass derived from broad H\(\alpha\) 
and broad H\(\beta\), to re-confirm the different kinematic properties (or different line profiles) between broad H$\alpha$ and broad 
H$\beta$ to support the central BSMBHs in \obj. Generally, the BH mass estimation by H\(\alpha\) is more reliable due to its 
lesser susceptibility to dust extinctions.

	Subsequently, the spatial separation between the two black holes can be further estimated based on total BH mass of a 
BSMBH system. The specific formula is as follows \citep{p94}:
\begin{equation}
A_{\text{BSMBH}} = 0.432 \times M_8 \times \left( \frac{P_{\text{BSMBH}} / \text{yr}}{2652 M_8} \right)^{2/3}
\end{equation}
with \(M_8 = \frac{M_{\text{BH}}}{10^8 M_\odot}\) and \( P_{\text{BSMBH}} \) as the orbital periodicity of the BSMBH 
system. Accepted the \( P_{\text{BSMBH}}\sim3.02\pm0.71\text{ yr} \)($1103\pm260$ days) and \( M(\text{BH,H}\alpha) \) as 
the total BH mass, the orbital separation can be estimated to be $A_{\text{BSMBH}}\sim(0.0053\pm0.0016)pc\sim6.3\pm1.9$ 
light-days. Meanwhile, based on the empirical R-L relation \(R_{\text{BLR}}\) $\propto$ $L_{5100}^{1/2}$ with the continuum luminosity 
$L_{5100}$=$(2.36\pm0.08)\times 10^{44}{\rm erg/s}$ at 5100\AA, the estimated \(R_{\text{BLR}}\) is $57\pm1$ light-days which is greater 
than the orbital separation $A_{BBH}$. It means that the two BLRs related to each BH accreting system in the supposed BSMBH 
system should be partly (or totally) mixed in \obj. Therefore, there are no apparent double-peaked features in the broad Balmer 
emission lines in \obj. Moreover, if the two BLRs were not totally mixed, periodic variability of the central wavelengths of the 
broad Balmer emission lines could be expected. In the future, using multi-epoch spectroscopic observations to search for velocity.

	Besides the preferred BSMBH system in \obj, the other explanations are discussed as follows for the optical QPOs 
in \obj, such as disk procession and jet processions.

	In \citet{as24}, large-scale radiation electromagnetic simulations were conducted, providing the first evidence that the 
procession phenomenon of super-strong accretion disks is driven by the rotation of the black hole. This procession leads to 
periodic variation in the direction of radiation of the accretion disk at different times. To explain the optical QPOs in \obj, 
the disc procession model discussed in \citet{el95} and \citet{bs03} has been accepted, leading to the expected periodicity as
\begin{equation}
T_{\text{pre}} \approx 1040 M_8 R_3^{2.5} \text{ yr} 
	\label{equation3}
\end{equation}
where \( R_3 \) (in units of ${\rm 10^3 R_G}$) represents the distance from the optical emission region to the central BH. 
Substituting the periodicity \( T \approx 3.02\pm0.71 \text{ yr} \) ($1103\pm260$ days), along with the BH mass 
\( M_(\text{BH, H}\alpha)\ \sim (1.13\pm0.14)\times 10^8 M_\odot \), the corresponding \( R_3 \) can be calculated to be 
0.092$\pm$0.0036, indicating the distance $R_{opt}$ about 92$\pm$3.6$R_{G}$ of the optical emission regions to central BH in \obj.

	Meanwhile, the distance of the near-ultraviolet (NUV) emission region to the central BH also can be estimated well by 
using the following formula \citep{mk10}:
\begin{equation}
\log R_{2500} = 15.78 + 0.80 \log \left( \frac{M_{\text{BH}}}{10^9 M_\odot} \right) 
\end{equation}
The calculation shows that the distance of the NUV emission regions to central BH in \obj~ is approximately \
\( (1.05\pm0.1) \times 10^{15} \text{ cm} \), which corresponds to about $R_{UV}\sim$\( 63\pm2 R_G \) 
(\( R_G = \frac{G M_{\text{BH}}}{C^2} \)). It is smaller than the $R_{opt}$ calculated by the disc procession model. Therefore, 
the disk procession model can not be rule out in \obj.

	However, if the disk procession was preferred in \obj, different periodicities in different epochs could be expected in \obj. 
For the ZTF light curves which are brighter than the CSS V-band light curve, simply indicating that the distance of optical emission 
regions to central BH during the period for ZTF g/r-band light curves should be longer than the distance during the period for the 
CSS V-band light curve, therefore larger periodicity should be expected in the ZTF g/r-band light curves. However, as shown in 
Figure~\ref{gls}, the periodicities in the ZTF light curves are smaller than that in the CSS light curve. Therefore, the disk 
procession is not preferred in \obj. In the near future, to check optical QPOs in different epochs could provide further clues to 
support or to rule out the disk procession in \obj.

	Furthermore, jet procession is one of the potential mechanisms for QPOs. Because the jet has a helical structure \citep{bh18}, 
there is a periodic change in the angle between its radiation direction and the observed line of sight, so that periodic variability 
can be observed, such as there is a periodicity about 47 days in 3C 454.3 was reported by \citet{sg21}. Like what has been done in 
the work of \citet{ks89}, the radio-optical luminosity ratio \( R \) can be used to identify the type of radio source, with the 
specific formula 
\begin{equation} 
R = \frac{\frac{f_{\text{r}} \times 5 \text{ GHz}}{10^{26} \text{ erg/s/Hz/cm}^2}}
	{\frac{f_{5100} \times 5100}{\text{erg/s/cm}^2/\text{\AA}}} 
\end{equation}
with \( f_{\text{r}} \)(3.19 mJy) the peak radio flux intensity at 5GHz, collected from the FIRST (Fain Images of the Radio Sky at 
Twenty-cm) \citep{bw95, hw15}(\href{http://sundog.stsci.edu/cgi-bin/searchfirst}{http://sundog.stsci.edu/cgi-bin/searchfirst}), and 
\( f_{5100} \) denoting the optical flux density at 5100\AA, $33.43\times10^{-17} \text{erg/s/cm}^2/\text{\AA}$. We can find 
R = $4.08 \times 10^{-2}$, indicating that \obj~ is a radio quiet object without significant jet activities. Thus, based on its 
radio properties, jet procession can be ruled out as the cause of the QPOs in \obj.

\subsection{Sub-pc triple BH system candidate related to QPOs with two periodicities}
	
	In the section above, as shown in Figure \ref{ztfg},\ref{wwz}, a comprehensive analysis of multi-band light curves, combined 
with cross-verification using multiple methods, confirmed the presence of QPOs in \obj\ with a periodicity about $1103 \pm 260$ days. 
However, besides the prominent periodicity around 1103 days, there is a secondary periodicity about $243 \pm 29$ days  
in the more refined ZTF light curve data.

	The ratio between these two periodicities is approximately $4.54\pm0.47$  showing an intriguing near-multiple relationship. 
This relationship may imply the existence of harmonic oscillations within the system, though limitations in current observational 
capabilities make it challenging to definitively identify which periodicity represents the true QPO signals.

	However, if both periodicities represent independent QPO phenomena, as there are few harmonic oscillations in the reported 
optical QPOs. In this case, the system might be a triple BH system, where gravitational interactions among the three black 
holes generate a complex dynamical structure. If assuming a triple BH system including a sub-pc BSMBH system and a 
third BH far away from the BSMBH system, some basic structure information can be simply estimated. Accepted the total 
BH mass of \( M(\text{BH,H}\alpha) \approx (1.13\pm0.14) \times 10^8 M_\odot \), the larger orbital separation corresponding to the 
$1103 \pm 260$ days periodicity is approximately $0.0053\pm0.0016$ pc. If accepted the total BH mass of the sub-pc BSMBH 
system and the third BH each account for half of the total BH mass, the smaller orbital separation associated with the $243\pm 29$ days 
periodicity is $0.00036\pm0.00004$ pc for the sub-pc close BH pair.

	Actually, despite the near-multiple characteristic observed between the two periods, limitations in current observational 
results make it difficult to clearly distinguish which of these periods represents the true QPO signal in the system.

\section{Summary and Conclusions}
The final summary and main conclusions are as follows. 
\begin{itemize}
\item A sinusoidal fit has been conducted on the light curves of CSS V-band and ZTF g/r-band of \obj, identifying a periodicity 
	about 985$\pm8$. The F-test technique can be applied to confirm the significance of the sinusoidal component with a 
	confidence level higher than $5\sigma$ (99.99994\%). Phase-folded light curves further supported the presence of optical QPOs.	
\item When analyzing ZTF g-band and r-band light curves independently, dual sinusoidal fits determined periodicity are $1024\pm12$ 
	days and $222\pm1$ days in the ZTF g-band light curve, $884\pm6$ days and $257\pm1$ in the ZTF r-band light curve.
\item According to GLS method, the periodicity in the CSS-V band light curve is $1250 \pm 112$ days, with a confidence level higher 
	than $5\sigma$. For the ZTF g-band light curve, there are two periodicities with confidence level higher than $5\sigma$, 
	$1165\pm 30$ days and $221\pm1.5$ days. And for the r-band light curve, there are also two periodicities with confidence 
	level higher than $5\sigma$, $1000\pm30$ days and $215\pm1$ days. The bootstrap methods further verifies the robustness of 
	periodic testing.	
\item	The WWZ method validates the periodicities of $1405 \pm 40$ days and $225 \pm 3$ days in the ZTF g-band, as well as 
	$950 \pm 42$ days and $270 \pm 3.5$ days in the r-band light curve. While a $1265\pm40$ days periodic signal in the CSS-V 
	band, the result is basically the same as those of the GLS method.
\item   The two periodicities in SDSS J1004+1510 are not related to intrinsic AGN activities, as confirmed by simulated 
	light curves using the CAR process, with a significance level higher than $3.17\sigma$.		
\item	Analysis of the H\(\alpha\) and H\(\beta\) broad emission line profiles of \obj~ yields a central virial BH mass of 
	\((1.13 \pm 0.14) \times 10^8 M_\odot\).	
\item	Based on the virial BH mass and optical periodicity, the estimated separation for two BHs
	is approximately \(A_{\text{BSMBH}} \sim 0.0053\pm0.0016 \text{ pc}\), if accepted BSMBH system in \obj.	
\item	Further analysis can lead the estimated size of the near-UV emitting region to be around \( 63\pm2 R_G \), it is one third times 
	smaller than the size of optical emitting regions (\(92\pm3.6 R_G\)) calculated according to the disk procession model. 
	It is hard to rule out the disk precession leading to the optical QPOs in \obj. However, due to smaller periodicities in 
	the ZTF light curves than that in the CSS light curve, the disk procession is not preferred in \obj.
\item   The peak radio flux of \obj~ is 1.39 mJy at 5GHz, leading \obj~ as a radio quiet AGN. Therefore, the jet procession can be 
	totally ruled out for the QPOs in \obj.
\item	For the source of multi-periodic QPO signals, two possibilities are proposed: one is that they may be harmonic signals 
	generated by base frequency oscillations. Another one is that there is a triple black hole system. Future observations will 
	provide further information to verify the origin of multi-periodic QPOs in \obj.
\end{itemize}

	\section*{Acknowledgements}
	The authors gratefully acknowledge the anonymous referee for giving us constructive comments and suggestions to greatly improve the paper. Zhang gratefully acknowledges the kind grant support 
	from NSFC-12173020 and NSFC-12373014. This paper has made use of the data from the SDSS projects, \url{http://www.sdss3.org/}, 
	managed by the Astrophysical Research Consortium for the Participating Institutions of the SDSS-III Collaboration. This 
	paper has made use of the data from the ZTF \url{https://www.ztf.caltech.edu} and from the CSS \url{http://nesssi.cacr.caltech.edu/DataRelease/}. 
	The paper has made use of the MPFIT package \url{https://pages.physics.wisc.edu/~craigm/idl/cmpfit.html}. This research has made use of the 
	NASA/IPAC Extragalactic Database (NED, \url{http://ned.ipac.caltech.edu}) which is operated by the California Institute of Technology, under 
	contract with the National Aeronautics and Space Administration.



\begin{thebibliography}{   }
	\bibitem[\protect\citeauthoryear{An, Lu \& Wang}{2016}]{an16} 
	An, T.; Lu, X.; Wang, J., 2016, A\&A, 585, 89
	\bibitem[\protect\citeauthoryear{Abraham}{2000}]{ab00}
	Abraham, Z., 2000, A\&A, 355, 915
	\bibitem[\protect\citeauthoryear{Ahumada et al.}{2021}]{ah21}
	Ahumada, R.; Prieto, C. A.; Almeida, A.; et al., 2021, ApJS, 249, 3
\bibitem[\protect\citeauthoryear{Asahina \& Ohsuga}{2024}]{as24}
	Asahina, Y.; Ohsuga, K., 2024, ApJ, 973, 45
	\bibitem[\protect\citeauthoryear{Bentz et al.}{2010}]{bk10}
	Bentz, M. C.; Kelly, D. D.; Catherine, J. G.; et al., 2010, ApJ 716 993
	\bibitem[\protect\citeauthoryear{Bundy et al.}{2009}]{bf09}
	Bundy, K.; Fukugita, M.; Ellis, R. S.; Targett, T. A.; Belli, S.; Kodama, T., 2009, ApJ, 697, 1369
	\bibitem[\protect\citeauthoryear{Bottrell et al.}{2019}]{bh19}
	Bottrell, C.; Hani, M. H.; Teimoorinia, H; et al., 2019, MNRAS, 490, 5390
\bibitem[\protect\citeauthoryear{Begelman, Blandford \& Rees}{1980}]{bb80}
	Begelman, M. C.; Blandford, R. D.; Rees, M. J., 1980, Natur, 287, 307
	\bibitem[\protect\citeauthoryear{Barth et al.}{2015}]{bb15}
	Barth, A. J.; Bennert, V. N.; Canalizo, G.; et al., 2015, ApJS, 217, 26
	\bibitem[\protect\citeauthoryear{Bergmann et al.}{2003}]{bs03}
	Bergmann, T. S.; Silva, R. N.; Eracleous, M.; et al., 2003, ApJ, 598, 956
	\bibitem[\protect\citeauthoryear{Bellm et al.}{2019}]{bk19}
	Bellm, E. C.; Kulkarni, S. R.; Barlow, T.; et al., 2019, PASP, 131, 068003
	\bibitem[\protect\citeauthoryear{Bretthorst}{2001}]{br01}
	Bretthorst, G. L., 2001, AIPC, 568, 241
	\bibitem[\protect\citeauthoryear{Becker, White \& Helfand}{1995}]{bw95}
	Becker, R. H.; White, R. L.; Helfand, D. J., 1995, ApJ, 450, 559
	\bibitem[\protect\citeauthoryear{Bentz et al.}{2013}]{bd13}
	Bentz, M. C.; Denney, K. D.; Grier, C. J.; et al., 2013, ApJ, 767, 149
	\bibitem[\protect\citeauthoryear{Bhatta}{2018}]{bh18}
	Bhatta, G., 2018, Galaxy, 6, 136
	\bibitem[\protect\citeauthoryear{Barnes \& Hernquist}{1996}]{bh96}
	Barnes, J. E.; Hernquist, L., 1996, ApJ, 471, 115
	\bibitem[\protect\citeauthoryear{Carlberg}{1992}]{ca92}
	Carlberg R. G., 1992, ApJL, 399, L31
\bibitem[\protect\citeauthoryear{Chen, Yu \& Lu}{2020}]{cy20}
		Chen, Y. F.; Yu, Q. J.; Lu, Y. J., 2020, ApJ, 897, 86
	\bibitem[\protect\citeauthoryear{Centrella et al.}{2010}]{cb10}
		Centrella J.; Baker, J. G.; Kelly, B. J.; van Meter, J. R., 2010, RvMP, 82, 3069
	\bibitem[\protect\citeauthoryear{Camenzind \& Krockenberger}{1992}]{ck92}
		Camenzind, M.; Krockenberger, M., 1992, A\&A, 255, 59
	\bibitem[\protect\citeauthoryear{Caproni, Abraham \& Monteiro}{2013}]{ca13}
		Caproni, A.; Abraham, Z.; Monteiro, H., 2013, MNRAS, 428, 280
	\bibitem[\protect\citeauthoryear{Charisi et al.}{2016}]{cb16}
		Charisi, M.; Bartos, I.; Haiman, Z.; Price-Whelan, A. M.; Graham, M. J.; Bellm, E. C.; Laher, R. R.; Marka, S., 2016, MNRAS, 463, 2145
	\bibitem[\protect\citeauthoryear{Drake et al.}{2009}]{dd09}
		Drake, A. J.; Djorgovski, S. G.; Mahabal, A.; et al., 2009, ApJ, 696, 870
	\bibitem[\protect\citeauthoryear{Dekany et al.}{2020}]{ds20}
		Dekany, R.; Smith, R. M.; Riddle, R.; et al, 2020, PASP, 132, 038001
		\bibitem[\protect\citeauthoryear{Eracleous et al.}{2012}]{p94}
		Eracleous M., Boroson T. A., Halpern J. P., Liu J., 2012, ApJS, 201, 23
	\bibitem[\protect\citeauthoryear{Eracleous, Livio \& Halpern}{1995}]{el95}
		Eracleous, M.; Livio, M.; Halpern, J. P., 1995, ApJ, 438, 610
	\bibitem[\protect\citeauthoryear{Ehlers et al.}{1976}]{er76}
		Ehlers, J.; Rosenblum, A.; Goldberg, J. N.; Havas, P., 1976, ApJL, 208, L77
	\bibitem[\protect\citeauthoryear{Foster}{1996}]{fo96}
		Foster, G., 1996, AJ, 112, 1709
	\bibitem[\protect\citeauthoryear{Ferrarese \& Ford }{2005}]{ff05}
		Ferrarese, L.; Ford, H., 2005, Space Sci. Rev., 116, 523
	\bibitem[\protect\citeauthoryear{Fragione et al.}{2019}]{fg19}
		Fragione, G.; Grishin, E.; Leigh, N. W. C.; Perets, H. B.; Perna, R., 2019, MNRAS, 488, 47
	\bibitem[\protect\citeauthoryear{Flanagan \& Cutler}{1994}]{fc94}
		Flanagan, E. E.; Cutler, C., 1994, PhRvD, 49, 2658
	\bibitem[\protect\citeauthoryear{Gaskell}{2010}]{ga10}
		Gaskell, C. M., 2010, Natur, 463, E1
	\bibitem[\protect\citeauthoryear{Graham et al.}{2015a}]{gd15a}
		Graham, M. J., Djorgovski, S. G., Stern, D.; et al., 2015a, Natur, 518, 74
	\bibitem[\protect\citeauthoryear{Graham et al.}{2015b}]{gd15b}
		Graham, M. J.; Djorgovski, S. G.; Stern, D.; et al., 2015b, MNRAS, 453, 1562	
	\bibitem[\protect\citeauthoryear{Gupta et al.}{2018}]{gt18}
		Gupta, A. C.; Tripathi, A.; Wiita, P. J.; et al., 2018, A\&A, 616, 6
	\bibitem[\protect\citeauthoryear{Greene \& Ho}{2005}]{gh05}
		Greene, J. E.; Ho, L. C.,2005, ApJ, 630, 122
	\bibitem[\protect\citeauthoryear{Heckman \& Best}{2014}]{hb14}
		Heckman, T. M.; Best, P. N., 2014, ARA\&A, 52, 589
	\bibitem[\protect\citeauthoryear{Hughes}{2021}]{hu21}
		Hughes, S. A., 2009, ARA\&A, 47, 107
	\bibitem[\protect\citeauthoryear{Huang et al.}{2021}]{hy21}
		Huang, S. F.; Yin, H. X.; Hu, S. M.; Xu, C.; jiang, Y. G.; Alexeeva, S.; Wang, Y. F., 2021, ApJ, 922, 222
	\bibitem[\protect\citeauthoryear{Helfand, White \& Becker}{2015}]{hw15}
		Helfand, D. J.; White, R. L.; Becker, R. H., 2015, ApJ, 801, 26
	\bibitem[\protect\citeauthoryear{Ingram et al.}{2016}]{iv16}
		Ingram, A.; van der Klis; M., Middleton, M.; et al., 2016, MNRAS, 461, 1967
	\bibitem[\protect\citeauthoryear{Jackson et al.}{2021}]{jk21}
		Jackson, R. A.; Kaviraj, S.; Martin, G.; et al., 2021, MNRAS, 506, 4499
	\bibitem[\protect\citeauthoryear{Korista \& Goad}{2004}]{kg04}
		Korista, K. T.; Goad, M. R., 2004, ApJ, 606, 749
	\bibitem[\protect\citeauthoryear{Kauffmann, White \& Guiderdoni}{1993}]{kw93}
		Kauffmann, G.; White, S. D. M.; Guiderdoni, B., 1993, MNRAS, 264, 201
	\bibitem[\protect\citeauthoryear{Kim et al.}{2024}]{kk24}
		Kim, D.; Kyeong, S. Y.; Jaffe, Y. L.; et al., 2024, ApJ, 966, 124
	\bibitem[\protect\citeauthoryear{Kormendy \& Richstone}{1995}]{kr95}
		Kormendy, J.; Richstone, D., 1995, ARA\&A, 33, 581	
	\bibitem[\protect\citeauthoryear{Kormendy \& Ho}{2013}]{kh13}
		Kormendy, J.; Ho, L. C., 2013, ARA\&A, 51, 511
	\bibitem[\protect\citeauthoryear{Kovacevic et al.}{2020}]{ky20}
		Kovacevic, A. B.; Yi, T. F.; Dai, X. Y.; Yang, X.; Hajdinjak, I. C.; Popovic, L. C., 2020, MNRAS, 494, 4069
	\bibitem[\protect\citeauthoryear{Kormendy et al.}{2009}]{kf09}
		Kormendy, J.; Fisher, D. B.; Cornell, M. E.; Bender, R., 2009, ApJS, 182, 216
	\bibitem[\protect\citeauthoryear{Kushwaha et al.}{2020}]{ks20}
		Kushwaha, P.; Sarkar, A.; Gupta, A. C.; Tripathi, A.; Wiita, P. J., 2020, MNRAS, 499, 653
	\bibitem[\protect\citeauthoryear{Kharb, Lal \& Merritt}{2017}]{kl17}
		Kharb, P.; Lal, D. V.; Merritt, D., 2017, NatAs, 1, 727
	\bibitem[\protect\citeauthoryear{Kaspi et al.}{2000}]{ks00}
		Kaspi, S.; Smith, P. S.; Netzer, H.; Maoz, D.; Jannuzi, B. T.; Giveon, U., 2000, ApJ, 533, 631
	\bibitem[\protect\citeauthoryear{Kellermann et al.}{1989}]{ks89}
		Kellermann, K. I.; Sramek, R.; Schmidt, M.; Shaffer, D. B.; Green, R., 1989, AJ, 98, 1195
	\bibitem[\protect\citeauthoryear{Kelly,Bechtold \&Siemiginowska}{2009}]{kbs09}
		Kell B. C.; Bechtold J.; Siemiginowska A., 2009, APJ, 698, 895
	\bibitem[\protect\citeauthoryear{Kozlowski et al.}{2010}]{k10}
		Kozlowski S. et al., 2010, APJ, 708, 927
	\bibitem[\protect\citeauthoryear{Li et al.}{2021}]{lc21}
		Li, X.; Cai, Y.; Yang, H.; Luo, Y.; Yan, Y.; He, J.; Wang, L., 2021, MNRAS,506, 2540
	\bibitem[\protect\citeauthoryear{Lacey \& Cole}{1994}]{lc94}
		Lacey, C.; Cole, S., 1994, MNRAS, 271, 676
	\bibitem[\protect\citeauthoryear{Lin et al.}{2004}]{lk04}
		Lin, L.; Koo, D. C.; Willmer, C. N. A.; et al., 2004, ApJL, 617, 9
	\bibitem[\protect\citeauthoryear{Lokas}{2023}]{le23}
		Lokas, E. L., 2023, A\&A, 673, A131
	\bibitem[\protect\citeauthoryear{Lauer \& Boroson}{2009}]{lb09}
		Lauer, T. R.;  Boroson, T. A., 2009, ApJ, 703, 930
	\bibitem[\protect\citeauthoryear{Liu et al.}{2015}]{lg15}
		Liu, T. T.; Gezari, S.; Heinis, S.; et al., 2015, ApJL, 803, L16
	\bibitem[\protect\citeauthoryear{Liu et al.}{2014}]{ls14}
		Liu, X.; Shen, Y.; Bian, F.; Loeb, A.; Tremaine, S., 2014, ApJ, 789 140
	\bibitem[\protect\citeauthoryear{Liao et al.}{2021}]{lc21}
		Liao, W. T.; Chen, Y. C.; Liu, X.; et al., 2021, MNRAS, 500, 4025
	\bibitem[\protect\citeauthoryear{Li et al.}{2016}]{lw16}
		Li, Y. R.; Wang, J. M.; He, Z. Q.; et al., 2016, ApJ, 822, 4
	\bibitem[\protect\citeauthoryear{Lomb}{1976}]{lo76}
		Lomb, N. R., 1976, Ap\&SS 39 447
	\bibitem[\protect\citeauthoryear{Mahabal et al.}{2011}]{md11}
		Mahabal, A. A.; Djorgovski, S. G.; Drake, A. J.; et al., 2011, BASI, 39, 387
	\bibitem[\protect\citeauthoryear{Menou, Haiman \& Narayanan}{2001}]{mh01}
		Menou, K.; Haiman, Z.; Narayanan, V. K., 2001, ApJ, 558, 535
	\bibitem[\protect\citeauthoryear{Merritt}{2006}]{md06}
		Merritt, D., 2006, ApJ, 648, 976
	\bibitem[\protect\citeauthoryear{Martin et al.}{2021}]{mj21}
		Martin, G.; Jackson, R. A.; Kaviraj, S., et al., 2021, MNRAS, 500, 4937
	\bibitem[\protect\citeauthoryear{Mayer et al.}{2010}]{mk10}
		Mayer, L.; Kazantzidis, S.; Escala, A.; Callegari, S., 2010, Natur, 466, 1082
	\bibitem[\protect\citeauthoryear{Mannerkoski et al.}{2022}]{mj22}
		Mannerkoski, M.; Johansson, P. H.; Rantala, A.; Naab, T.; Liao, S.; Rawlings, A., 2022, ApJ, 929, 167
	\bibitem[\protect\citeauthoryear{Merritt \& Milosavljevic}{2005}]{mm05}
		Merritt, D.; Milosavljevic, M., 2005, Living Reviews in Relativity, 8, 8
	\bibitem[\protect\citeauthoryear{ Marscher \& Gear}{1985}]{mg85}
		Marscher, A. P.; Gear, W. K., 1985, ApJ, 298, 114
	\bibitem[\protect\citeauthoryear{Morgan et al.}{2010}]{mk10}
		Morgan, C. W.; Kochanek, C. S.; Morgan, N. D.; Falco, E. E., 2010, ApJ, 712, 1129
	\bibitem[\protect\citeauthoryear{MacLeod et al.}{2010}]{m10}
		MacLeod C. L., et al., 2010, APJ, 721, 1014
	\bibitem[\protect\citeauthoryear{Netzer}{2020}]{nh20}
		Netzer, H., 2020, MNRAS, 494, 1611
	\bibitem[\protect\citeauthoryear{O'Neill et al.}{2022}]{ok22}
		O'Neill, S.; Kiehlmann, S.; Readhead, A. C. S.; et al., 2022, ApJL, 926, L35
	\bibitem[\protect\citeauthoryear{Pihajoki, Valtonen \& Ciprini}{2013}]{pv13}
		Pihajoki, P.; Valtonen, M.; Ciprini, S., 2013, MNRAS, 434, 3122
	\bibitem[\protect\citeauthoryear{Peterson et al.}{2004}]{pe04}
		Peterson, B. M.; Ferrarese, L.; Gilbert, K. M.; et al., 2004, ApJ, 613, 682
	\bibitem[\protect\citeauthoryear{Rodriguez et al.}{2016}]{rp16}
		Rodriguez, G, V.; Pillepich, A.; Sales, L. V.; et al., 2016, MNRAS, 458,2371
	\bibitem[\protect\citeauthoryear{Rodriguez et al.}{2017}]{rs17}
		Rodriguez, G, V.; Sales, L. V.; Genel, S.; et al., 2017, MNRAS, 467,3083
	\bibitem[\protect\citeauthoryear{Rodriguez et al.}{2006}]{rt06}
		Rodriguez, C.; Taylor, G. B.; Zavala, R. T.; Peck, A. B.; Pollack, L. K.; Romani, R. W., 2006, ApJ, 646, 49
	\bibitem[\protect\citeauthoryear{Runnoe et al.}{2015}]{re15}
		Runnoe, J. C.; Eracleous, M., Mathes, G.; et al., 2015, ApJS, 221, 7
		\bibitem[\protect\citeauthoryear{Silk \& Rees}{1998}]{sr98}
		Silk, J.; Rees, M. J., 1998, A\&A, 331, L1	
	\bibitem[\protect\citeauthoryear{Satyapal et al.}{2014}]{se14}
		Satyapal, S.; Ellison, S. L.; McAlpine, W.; Hickox, R. C.; Patton, D. R.; Mendel, J. T., 2014, MNRAS, 441, 1297
	\bibitem[\protect\citeauthoryear{Shields et al.}{2009}]{ss09}
		Shields, G. A.; Smith, K. L.; Salviander, S.; Strickler, R.; Dutton, A. A.; Marshall, P. J., 2009, ApJ, 707, 936
	\bibitem[\protect\citeauthoryear{Shen et al.}{2013}]{sl13}
		Shen Y., Liu X., Loeb A., Tremaine S., 2013, ApJ, 775, 49
	\bibitem[\protect\citeauthoryear{Shu et al.}{2020}]{sz20}
		Shu, X.; Zhang, W.; Li, S.; et al., 2020, NatCo, 11 5876
	\bibitem[\protect\citeauthoryear{Stella \& Vietri}{1998}]{sv98}
		Stella, L.; Vietri, M., 1998, ApJL, 491, L59
	\bibitem[\protect\citeauthoryear{Stella, Vietri \& Morsink}{1999}]{sv99}
		Stella, L.; Vietri, M.; Morsink, S. M., 1999, ApJL, 524, L63
	\bibitem[\protect\citeauthoryear{Songsheng et al.}{2020}]{sx20}
		Songsheng, Y. Y.; Xiao, M.; Wang, J. M.; Ho, L. C., 2020, ApJS, 247, 3
	\bibitem[\protect\citeauthoryear{Sesana et al.}{2018}]{sh18}
		Sesana, A.; Haiman, Z.; Kocsis, B.; Kelley, L. Z., 2018, ApJ, 856, 42
	\bibitem[\protect\citeauthoryear{Serafinelli et al.}{2020}]{ss20}
		Serafinelli, R.; Severgnini, P.; Braito, V.; et al., 2020, ApJ, 902, 10
	\bibitem[\protect\citeauthoryear{Scargle}{1982}]{sc82}
		Scargle J. D., 1982 ApJ 263 835
	\bibitem[\protect\citeauthoryear{Springford, Eadie \& Thomson}{2020}]{se20}
		Springford, A.; Eadie, G. M.; Thomson, D. J., 2020, AJ, 159, 205	
	\bibitem[\protect\citeauthoryear{Sarkar et al.}{2021}]{sg21}
		Sarkar, A.; Gupta, A. C.; Chitnis, V. R., et al., 2021, MNRAS, 501, 50
	\bibitem[\protect\citeauthoryear{Tsalmantza et al.}{2011}]{td11}
		Tsalmantza, P.; Decarli, R.; Dotti, M.; Hogg, D. W., 2011, ApJ, 738, 20
	\bibitem[\protect\citeauthoryear{Tsang \& Lai}{2009}]{tl09}
		Tsang, D.; Lai, D., 2009, MNRAS, 396, 589
	\bibitem[\protect\citeauthoryear{Vaughan et al.}{2016}]{vu16}
		Vaughan, S.; Uttley, P.; Markowitz, A. G.; et al., 2016, MNRAS, 461, 3145
	\bibitem[\protect\citeauthoryear{Vietri \& Stella}{1998}]{vs98}
		Vietri, M.; Stella, L. G., 1998, ApJ, 503, 350
	\bibitem[\protect\citeauthoryear{VanderPlas}{2018}]{vt18}
		VanderPlas, J. T., 2018, ApJS, 236, 16
	\bibitem[\protect\citeauthoryear{Vestergaard \& Peterson}{2006}]{vp06}
		Vestergaard, M.; Peterson, B. M. 2006, ApJ, 641, 689
	\bibitem[\protect\citeauthoryear{Wang et al.}{2017}]{wg17}
		Wang, L. L.; Greene, J. E.; Ju, W. H.; Rafikov, R. R.; Ruan, J. J.; Schneider, D. P., 2017, ApJ, 834, 129
	\bibitem[\protect\citeauthoryear{Wandel, Peterson \& Malkan}{1999}]{wp99}
		Wandel, A.; Peterson, B. M.; Malkan, M. A., 1999, ApJ, 526, 579
	\bibitem[\protect\citeauthoryear{Yoon et al.}{2022}]{yp22}
		Yoon, Y.; Park, C.; Chung, H.; Lane, R. R., 2022, ApJ, 925, 168
	\bibitem[\protect\citeauthoryear{Yang et al.}{2019}]{yg19}
		Yang, C.; Ge J.; Lu, Y., 2019, SCPMA, 62, 129511
	\bibitem[\protect\citeauthoryear{Zheng et al.}{2016}]{zb16}
		Zheng, Z. Y.; Butler, N. R.; Shen, Y.; Jiang, L. H.; Wang, J. X.; Chen, X.; Cuadra, J.,  2016, ApJ, 827, 56
	\bibitem[\protect\citeauthoryear{Zechmeister \& Kurster}{2009}]{zk09}
		Zechmeister, M.; Kurster, M., 2009, A\&A, 496, 577
	\bibitem[\protect\citeauthoryear{Zhang}{2023a}]{zh23a}
		Zhang X. G, 2023a, MNRAS, 525, 335.
	\bibitem[\protect\citeauthoryear{Zhang}{2022a}]{zh22a} 
		Zhang X. G.,2022a, MNRAS,512, 1003
	\bibitem[\protect\citeauthoryear{Zhang}{2023b}]{zh23b} 
		Zhang X. G., 2023b, MNRAS, 526, 1588.
	\bibitem[\protect\citeauthoryear{Zhang}{2022b}]{zh22b}
		Zhang X. G., 2022b, MNRAS, 516, 3650.
	\bibitem[\protect\citeauthoryear{Zhang}{2025a}]{zh25}
		Zhang X. G., 2025a, ApJ in press, arXiv:2412.15506
	\bibitem[\protect\citeauthoryear{Zhang}{2025b}]{zh25b}
		Zhang X. G., 2025b, ApJ in press, arXiv:2503.10050
	\bibitem[\protect\citeauthoryear{Zu et al.}{2013}]{z13}
		Zu Y.; Kochanek C. S.; Kozlowski S.; Udalski A., 2013, APJ, 765,106	
		
	\end{thebibliography}
\end{document}